\documentclass[conference]{IEEEtran}
\usepackage{url}
\usepackage{stfloats}
\usepackage{fixltx2e}
\usepackage[pdftex]{graphicx}

\usepackage{cite}
\usepackage{amsmath,amssymb,amsfonts}
\usepackage{algorithmic}
\usepackage{graphicx}
\usepackage{subfigure}
\usepackage{textcomp}
\usepackage{xcolor}
\usepackage{amsmath,amssymb,amsfonts}
\usepackage{algorithm}
\usepackage{threeparttable}
\usepackage{ragged2e}
\usepackage{listings}
\usepackage{booktabs,makecell, multirow, tabularx}
\usepackage{pifont}
\usepackage{adjustbox}
 \usepackage{hyperref} 
 \usepackage[misc]{ifsym}
\usepackage{xcolor}
\lstdefinestyle{mystyle}{
	moredelim=**[is][\color{red!100}]{@}{@},
}

\def\BibTeX{{\rm B\kern-.05em{\sc i\kern-.025em b}\kern-.08em
    T\kern-.1667em\lower.7ex\hbox{E}\kern-.125emX}}

\begin{document}

\title{DeepGo: Predictive Directed Greybox Fuzzing}

\author{\IEEEauthorblockN{Peihong Lin, Pengfei Wang, Xu Zhou, Wei Xie, Gen Zhang, Kai Lu}
				\IEEEauthorblockA{College of Computer, National University of Defense Technology\\
				\{phlin22, pfwang, zhouxu, xiewei, zhanggen, kailu\}@nudt.edu.cn}
                }

\IEEEoverridecommandlockouts
\makeatletter\def\@IEEEpubidpullup{6.5\baselineskip}\makeatother
\IEEEoverridecommandlockouts
\makeatletter\def\@IEEEpubidpullup{6.5\baselineskip}\makeatother
\IEEEpubid{\parbox{\columnwidth}{
    Network and Distributed System Security (NDSS) Symposium 2024\\
    26 February - 1 March 2024, San Diego, CA, USA\\
    ISBN 1-891562-93-2\\
    https://dx.doi.org/10.14722/ndss.2024.24514.\\
    www.ndss-symposium.org
}
\hspace{\columnsep}\makebox[\columnwidth]{}}

\maketitle

\renewcommand{\thefootnote}{\fnsymbol{footnote}}
\footnotetext[1]{These authors contributed equally to this work.}

\begin{abstract}
Directed Greybox Fuzzing (DGF) is an effective approach designed to strengthen testing vulnerable code areas via predefined target sites. 
The state-of-the-art DGF techniques redefine and optimize the fitness metric to reach the target sites precisely and quickly. However, optimizations for fitness metrics are mainly based on heuristic algorithms, which usually rely on historical execution information and lack foresight on paths that have not been exercised yet. Thus, those hard-to-execute paths with complex constraints would hinder DGF from reaching the targets, making DGF less efficient.

In this paper, we propose DeepGo, a predictive directed greybox fuzzer that can combine historical and predicted information to steer DGF to reach the target site via an optimal path. 
We first propose the \textit{path transition model}, which models DGF as a process of reaching the target site through specific path transition sequences. 
The new seed generated by mutation would cause the path transition, and the path corresponding to the high-reward path transition sequence indicates a high likelihood of reaching the target site through it.
Then, to predict the path transitions and the corresponding rewards,
we use deep neural networks to construct a Virtual Ensemble Environment (VEE), which gradually imitates the path transition model and predicts the rewards of path transitions that have not been taken yet.
To determine the optimal path, we develop a Reinforcement Learning for Fuzzing (RLF) model to generate the transition sequences with the highest sequence rewards. The RLF model can combine historical and predicted path transitions to generate the optimal path transition sequences, along with the policy to guide the mutation strategy of fuzzing. 
Finally, to exercise the high-reward path transition sequence, we propose the concept of an \textit{action group}, which comprehensively optimizes the critical steps of fuzzing to realize the optimal path to reach the target efficiently.
We evaluated DeepGo on 2 benchmark suites consisting of 25 programs with a total of 100 target sites. The experimental results show that DeepGo achieves 3.23$\times$, 1.72$\times$, 1.81$\times$, and 4.83$\times$ speedup compared to AFLGo, BEACON, WindRanger, and ParmeSan, respectively in reaching target sites, and 2.61$\times$, 3.32$\times$, 2.43$\times$ and 2.53$\times$ speedup in exposing known vulnerabilities.

\end{abstract}


\section{Introduction}
Directed Greybox Fuzzing (DGF) \cite{14} is an efficient technique designed for testing vulnerable code areas.
By defining a measurable fitness metric, the directed greybox fuzzer can select promising seeds and give them more mutation chances to approach the target site gradually.
For example, based on the call graph and control-flow graph information of the program under test (PUT), mainstream DGF techniques use the distance between inputs and target sites as the fitness metric to assist seed selection and seed energy assignment. Seeds that are closer to the target site are regarded as promising and prioritized.
DGF spends most of its time on reaching these locations without wasting resources stressing unrelated parts, thus, it is particularly suitable for testing scenarios including patch testing \cite{15}, bug reproduction \cite{18}, and potential buggy code verification \cite{19}.

To enhance directedness and improve efficiency, the state-of-the-art DGF works leverage heuristic methods to redefine and optimize fitness metrics. 
For instance, some DGF techniques redefine the fitness metric based on trace similarity (e.g., Hawkeye \cite{21}),  data conditions (e.g., CAFL \cite{22}), and data flow information (e.g., WindRanger \cite{23}). 
Some directed fuzzers use the sequence-based approach to enhance directedness, such as extending the given target sequence (e.g., Berry \cite{berry}) and taking the use-after-free sequence as guidance (e.g., Lolly \cite{lolly}).
And some DGF techniques improve efficiency by pruning unreachable paths to the target (e.g., BEACON \cite{26}) and constructing the queryable oracle to guide fuzz ($MC^2$ \cite{MC}). 
However, in fuzzing, the mutation of seeds makes the execution path uncertain. Those heuristic methods usually rely on historical execution information and lack foresight on paths that have not been exercised yet.
For example, when using the basic block level distance to the target as the fitness metric, seeds with a shorter distance are prioritized without considering the path feasibility, as a result, those hard-to-execute paths with complex constraints would hinder DGF from reaching the target sites, making DGF less efficient. 
Therefore, in this paper, we aim to design a predicted directed greybox fuzzer that can foresee critical execution information and predict the \textit{optimal path}. By combining the historical execution information and the predicted future execution information, the fuzzer can intelligently generate the optimal and viable path to the target site. By avoiding the infeasible and hard-to-execute paths, the fuzzer can reach the target site more precisely and efficiently.

For this purpose, we propose to \textit{model DGF as a process of reaching the target site through specific path transition sequences} and name the model as the \textbf{\textit{path transition model}}. 
The new seed generated by mutation would cause the path transition, and we use \textit{reward} to evaluate the immediate impact of path transitions on the fuzzer. We use \textit{sequence reward} to evaluate the difficulty of reaching the target site through a sequence of path transitions. 
The path corresponding to the high-reward path transition sequence indicates a high likelihood of reaching the target site through it.
Compared to previous heuristic methods, this model takes the difficulty of reaching the target site through different path transition sequences into consideration. Besides, 
by analyzing the sequence rewards of different path transition sequences, an optimal path with the highest sequence reward can be generated, which can be used to guide the fuzzer to the target site more efficiently.
To achieve this goal, we need to address three challenges.

\textbf{Challenge 1: How to predict path transitions that have not been taken?} To generate the optimal path to the target site, we need to collect all the potential path transitions. However, the existing techniques can only collect information on the known path transitions, not suitable for future path transitions. In DGF, different mutations 
would cause different path transitions. Within the limited time budget, the fuzzer cannot try all mutations (e.g., using all mutators to mutate all seed bytes) on the seeds.
Besides, randomly selecting mutators and bytes for mutation can result in inefficient mutators being chosen and key bytes being missed.
Therefore, we should predict the potential path transitions and the corresponding rewards caused by the mutations that have not been taken.

\textbf{Challenge 2: How to determine the optimal path among large numbers of path transitions?} In the path transition model, a higher sequence reward indicates a higher likelihood of reaching the target site through the path transition sequence. The optimal path can be represented by a path transition sequence with the highest sequence reward. However, the combination of historical and predicted path transitions creates large numbers of path transition sequences with different sequence rewards, Thus, we should efficiently evaluate the sequence rewards of all path transition sequences and design a policy to guide the mutations to realize the optimal path. 

\textbf{Challenge 3: How to exercise the optimal path transition sequences by optimizing the fuzzing strategies?} Since DGF still employs random mutation strategies, the paths covered by the fuzzer are random and constantly changing. However, based on the path transition model, to efficiently steer the fuzzer towards the target site via the optimal path transition sequence, we should comprehensively optimize the critical steps of fuzzing, such as seed selection, energy assignment, mutator schedule, looping cycles in havoc \cite{rule}, and mutation location selection.
Therefore, we should be able to optimize the critical steps of fuzzing simultaneously to exercise the optimal path transition sequences efficiently.

To address these challenges, we propose DeepGo, a predictive directed greybox fuzzer. Based on the path transition model, we extract the path covered by one seed, mutation on the seed, path transition, and seed value changes caused by path transition in the DGF to a four-tuple \textit{(path, action, next\_path, reward)} (see Section \ref{ptm}).
 
For \textbf{Challenge 1}, we use deep neural networks (DNNs) to construct a Virtual Ensemble Environment (VEE). Given a path and an action,
the well-trained VEE can predict potential path transitions and the corresponding rewards.
With the increasing path transition information provided to VEE, VEE can gradually imitate the path transition model and predict the path transitions 
caused by the mutations that have not been taken, which greatly improves the efficiency of DGF. 
For \textbf{Challenge 2}, we propose a Reinforcement Learning for Fuzzing (RLF) model to determine the optimal path. To give RLF foresight, we use a k-step branch rollout strategy to continuously obtain predicted path transitions from VEE. 
By combining the historical and predicted path transitions, the RLF model is trained to evaluate the expected sequence rewards of path transition sequences caused by different mutations and determine the highest sequence reward. Additionally, the model can learn a policy for the optimal path to guide the mutation strategy of  fuzzing. 
For \textbf{Challenge 3}, to exercise the optimal path, we optimize fuzzing strategies based on the concept of \textit{action group}.  In the action group, we comprehensively consider the five critical steps, including seed selection, energy assignment, the selection of looping cycles in havoc \cite{rule}, mutator schedule, and mutation location selection.
Under the mutation policy generated by the RLF model, we use a Multi-elements Particle Swarm Optimization (MPSO) algorithm to optimize them simultaneously, to realize the desired mutations and generate the path transition sequences, and ultimately to reach the target site via the optimal path.

In summary, we mainly make the following contributions:

\begin{itemize}
	
\itemsep -0.1em
 
\item  We propose the path transition model, which models DGF as a process of reaching the target site through specific path transition sequences. Based on the path transition model, we use sequence reward as the fitness metric to evaluate the difficulty of reaching the target site through a sequence of path transitions.

\itemsep -0.1em 

\item We construct the VEE, using DNNs to imitate the path transition model and predict potential path transitions and the corresponding rewards without exercising the path, which greatly improves efficiency.

\itemsep -0.1em 

\item We propose the RLF model, which can combine the historical and predicted  information to generate the optimal path to the target site. By avoiding the infeasible and hard-to-execute path, the optimal path with the highest sequence reward can guide the fuzzer to the target site efficiently and precisely.

\item We optimize the mutation strategy of fuzzing at the granularity of the action group, which is more efficient than single-strategy optimization. In the action group, we consider the five critical steps of fuzzing and optimize them simultaneously with an MPSO algorithm.

\item We implement and evaluate DeepGo. The evaluation results demonstrate that the VEE can predict the path transitions with high accuracy and DeepGo can reach the target sites faster than baseline fuzzers. 

\itemsep -0.1em 
	
\item The artifact of DeepGo is available on our website. \url{https://gitee.com/paynelin/DeepGo}.

\itemsep -0.1em 

\end{itemize}

\section{Background}
\label{background}

\textbf{Directed Greybox Fuzzing}. Following AFLGo \cite{43}, the existing DGF techniques calculate the distances between the inputs and predefined targets based on the call graph and control-flow graph and combine the distance with other indicators (e.g., condition complexity) to form the fitness metric. Then, at runtime, the directed greybox fuzzing techniques design different power schedules according to the fitness metric to assign more energy to the seeds that are preferred. It casts reachability as an optimization problem to minimize the distance between the generated inputs and the targets. 

\textbf{Deep Neural Networks}. In recent years, Deep Neural Networks (DNNs) have demonstrated their ability to approximate complex non-linear and non-convex functions and imitate the environment in pattern recognition \cite{Deep1, Deep2}.  DNNs have been used in some fuzzing works, such as NEUZZ \cite{NEUZZ}, MTFuzz \cite{MTFUZZ} and FUZZGUARD \cite{27}, to simulate program branching behavior. In summary, these works collect the mutated inputs generated during the fuzzing process as the input of DNNs and record the covered program branches as the labels to train DNNs. By this means, the DNNs can simulate the program branching behavior and guide the fuzzing optimization. The evaluations of these works prove that DNNs are appropriate for simulating the fuzzing environment.

\textbf{Reinforcement Learning}. Reinforcement Learning (i.e., RL) \cite{DQN, D2QN, Dueling} is commonly used to solve sequential decision problems (such as the Markov process). In the RL model, the agents would constantly take actions to interact with the environment and receive feedback, namely rewards. Based on the feedback, represented as a four-tuple ($state, action, next\_state, reward$), from the environment, the RL model would optimize the action selection strategies (i.e., \textit{policy} in RL) to obtain the maximum rewards. Following AFLGo, we can model DGF as the Markov process and apply the RL model to optimize the fuzzing strategies, such as the selection of mutation operators and mutation bytes.

\textbf{Model-Based Policy Optimization \cite{MBPO}}. Model-Based Policy Optimization (MBPO) is a model-based reinforcement learning method consisting of two modules: a virtual environment and a reinforcement learning network. Firstly, MBPO uses DNNs to create a virtual environment to replace the real environment. Secondly, MBPO allows the agent to interact with the virtual environment to obtain a large number of path transitions. Then, MBPO can continuously train the reinforcement learning network and learn the policy that maximizes rewards for the agent based on the path transitions. In this paper, we refer to some methods from MBPO, such as the k-step branch rollout strategy, to combine DNNs and RL.
 
\textbf{Particle Swarm Optimization}. Particle Swarm Optimization (PSO) algorithm \cite{PSO} is an evolutionary computation technique that originated from the study of bird flocking behavior during foraging. It has been applied to improve the fuzzing efficiency in CGF (e.g.,  MOPT \cite{MOPT} uses the PSO algorithm to optimize the selection probability of mutation operators based on the fuzzing historical information). 
The main idea behind PSO is to find the optimal solution through collaboration and information sharing among individuals in the population. 
PSO algorithm uses massless particles to simulate birds in a flock, and these particles have two primary attributes: velocity and position. Velocity represents the speed of movement, and position represents the direction of movement. Each particle searches for the optimal solution individually in the search space and records it as the local best value. The local best values are then shared among the particles in the entire swarm to find the global best value as the optimal solution. All particles in the swarm would adjust their velocity and position based on the local best value and the global best value shared in the entire swarm.

\section{PATH TRANSITION MODEL}
\label{ptm}

In this paper, we propose the path transition model, which models DGF as a process of reaching the target site through specific path transition sequences. The new seeds generated by mutations would cause path
transitions, and we use rewards to evaluate the immediate impact
of path transitions on the fuzzer. The path transition sequence with the highest sequence reward determines the optimal path to the target.
In this section, we map the key elements in DGF to the path transition model and quantify the effectiveness of path transitions and actions.

\subsection{Elements in Path Transition Model}

\textbf{Path}. Each path corresponds to a seed in the seed queue of the fuzzer. We use \texttt{trace\_bits} in AFL \cite{43} to record the covered branches and branches' hit counts in the path and distinguish different paths. 

\textbf{Action}. A fuzzer's action means to mutate a seed at a specific location. We pay attention to the location (i.e., bytes) where the mutation occurs instead of the mutator it uses. The fuzzer takes a series of actions to reach the target site gradually.

\textbf{Path transition}. Mutation on the seed causes the path transition if the execution path of the new input is different from that of the seed. If the new input's path is the same as the seed's, the mutation causes a self-path-transition.

\textbf{Reward}. The reward for a path transition represents the change in the seed's value caused by the path transition. 

\textbf{Policy}. The policy is the strategy for the fuzzer to select actions in each path, represented as a list of probabilities corresponding to the actions. Under the policy, the fuzzer would select actions with different probabilities.

\begin{figure}[t]
	\setlength{\abovecaptionskip}{-0.5em} 
	\centering
	\noindent \includegraphics[width=7cm, height=4.5cm]{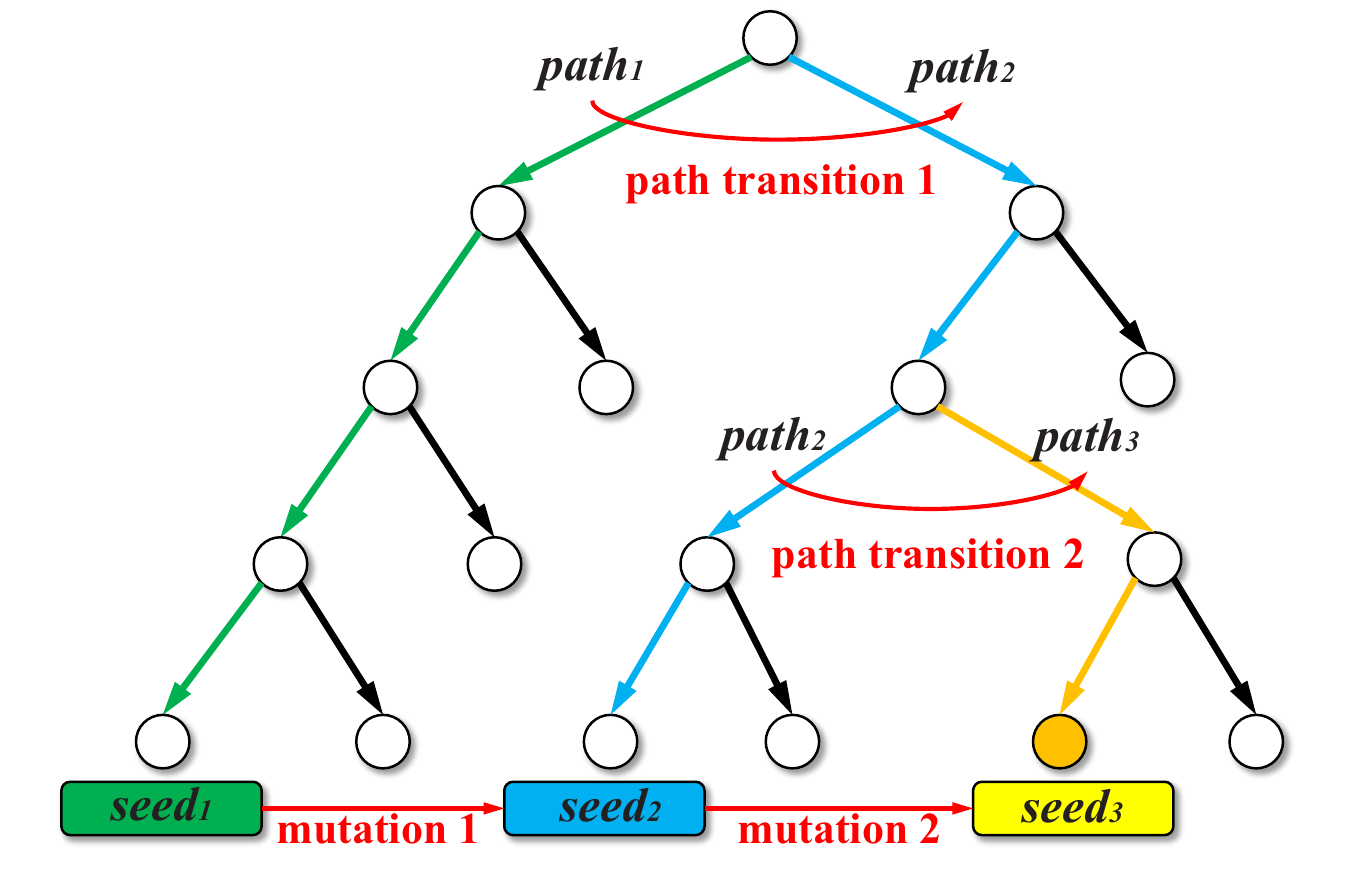}
	\caption{\label{fig:0} Illustration of the path transition model.} 
	\vspace{-0.5em} 
\end{figure}
We use Fig. \ref{fig:0} to illustrate the path transition model. Fig. \ref{fig:0} shows the execution tree of an example program, where the nodes represent the basic blocks and the edges represent the basic block transition. The execution path $path_1$ marked green is covered by seed $s_1$, the execution path $path_2$ marked blue is covered by seed $s_2$, and the execution path $path_3$ marked yellow is covered by seed $s_3$. The node marked as orange is the target basic block, and $path_3$ is the optimal path that can reach the target site. 
During the fuzzing process, firstly, the fuzzer takes action to mutate seed $s_1$ to generate a new seed $s_2$, making $path_1$ transfers to $path_2$. Then, the fuzzer takes action to mutate seed $s_2$, making $path_2$ transfers to $path_3$ and reaching the target site.
Through the two path transitions, the fuzzer generates the optimal path $path_3$ represented by a path transition sequence (\textit{path transition 1}, \textit{path transition 2}) that reaches the target.

\subsection{Quantifying path transitions and actions}
\label{value}

The fuzzer taking different actions to mutate seeds will cause different path transitions, we use \textit{rewards} to quantify the effectiveness of path transitions and use \textit{expected sequence rewards} to quantify the effectiveness of actions. 

\textbf{Seed value.} 
We use seed value to evaluate different paths based on their contribution to the fuzzer reaching the target site.
The seed value is calculated based on four characteristics of the seed corresponding to the path: (1) seed distance to the target, (2) the difficulty of satisfying the branch inversion, (3) execution speed, and (4) whether the seed is ``favored''. 
In DGF, a shorter seed distance implies that the fuzzer can reach the target site by satisfying fewer path constraints, and the lower difficulty implies that it is easier for the fuzzer to satisfy the path constraints leading to the target site. Moreover, during the seed execution process, the fuzzer may get trapped in loops (e.g., \texttt{while} and \texttt{for} loops), which reduces the execution speed and does not contribute to the fuzzer reaching the target site. Therefore, to enhance the fuzzing efficiency, we prefer seeds that have faster execution speed. Additionally, we prefer seeds that are marked as ``favored'' since these seeds can cover all the explored branches. By fuzzing the favored seeds, we can perform branch inversion on all explored branches to cover new branches leading to the target site. Following the methods of AFL \cite{5} and AFLGo \cite{43}, elements of (1), (3), and (4) can be obtained by simply recording the execution information during fuzzing. Here, we mainly explain the concept of ``difficulty'' and its calculation. 

We use branch probability to measure the difficulty of satisfying branch inversion, which is based on the statistics of branch hits.
We first obtained information on the sibling branches of each branch during the static analysis.
If the sibling branches of the covered branches are still uncovered (i.e., unexplored branches), satisfying the branch inversion to cover the unexplored branches would allow the fuzzer to transfer from the covered paths to new paths. 
Then, we count branch hits to calculate the branch probability of the unexplored branch. If the fuzzer consistently hits the covered branches with mutated inputs but cannot hit the unexplored branches, it indicates that the fuzzer has difficulties in satisfying the branch inversion.
Finally, we quantify the difficulty of satisfying the branch inversion by branch probability:
\begin{equation}
	\label{eq1}
	P(ubr) = {1 \over {\sum\limits_{br \in \phi (cond)} {hi{t_{br}} + 1} }}
\end{equation}
Where $ubr$ denotes an unexplored branch. We check whether the siblings of the covered branches are covered to find the unexplored branches. $\phi (cond)$ denotes the set of all branches under the same condition. $hi{t_{br}}$ denotes the branch hits recorded during fuzzing. $P(ubr)$ denotes the branch probability of the unexplored branches, which will not be 0 since we believe the unexplored branches always have the probability of being covered by fuzzing. We use the arithmetic mean of the branch probabilities of all unexplored branches in one seed's path to estimate the difficulty of satisfying the branch inversion:
\begin{equation}
	\label{eq2}
	ED{_s} = {{\sum\nolimits_{br \in \Phi (s)} {P(br)} } \over {|\Phi (s)|}}
\end{equation}
Where $s$ denotes the seed, $ED{_s}$ denotes the estimated difficulty, and ${\Phi (s)}$ denotes the set containing all the unexplored branches in the seed's path. 

Therefore, $distance$, $difficulty$, $execution \ speed$, and $favored$ can all be quantitatively measured and calculated to calculate the seed value. We use the Entropy Weight Method \cite{EWM} to determine the weights of the four factors based on their information entropy. A small information entropy value makes a small factor weight, indicating the factor has a small impact on the overall evaluation of the seed value.
\begin{equation}
\label{eq0}
{V^s}({p_t}) = {W_1} \cdot {d_s} + {W_2} \cdot E{D_s} + {W_3} \cdot E{x_s} + {W_4} \cdot F{v_s}
\end{equation}
Where ${V^s}({p_t})$ denotes the seed value of path $p_t$, $W_1$, $W_2$, $W_3$, and $W_4$ are the weights calculated based on the Entropy Weight Method. 
$d_s$ denotes the seed distance, $Ex_s$ denotes the execution speed, and $Fv_s$ indicates where it is $favored$, the value which could be 0 or 1. 
The Entropy Weight Method \cite{EWM} is a commonly used multi-element comprehensive evaluation method. The steps for calculating the weights of different elements are as follows:

\noindent (1) Calculate the entropy of each element:
\begin{equation}
	\label{eq21}
	E_j = -\frac{1}{\ln(n)}\sum_{i=1}^n \frac{p_{ij}}{\ln(p_{ij})}
\end{equation}
(2) Calculate the weight of each element:
\begin{equation}
	\label{eq22}
	W_j = \frac{1-E_j}{n-\sum_{j=1}^nE_j}
\end{equation}
(3) Normalize the weight of each element:
\begin{equation}
	\label{eq23}
	W_j' = \frac{W_j}{\sum_{j=1}^n w_j}
\end{equation}
Where $W_j'$ denotes  the normalized weight of the $j^{th}$ element. The core idea of the Entropy Weight Method is to determine the weight of each element by calculating their entropies, so as to conduct multi-element comprehensive evaluation.

Then, we calculate the reward based on the seed value to evaluate the effectiveness of the path transition:
\begin{equation}
	r({p_t},{a_t}, p_{t+1}) = {V^s}({p_{t + 1}}) - {V^s}({p_t})
	\label{eq5}
\end{equation}
Where $r({p_t},{a_t},p_{t+1})$ denotes the reward, $a_t$ denotes the action that the fuzzer takes to transfer from path $p_{t}$ to path $p_{t+1}$, ${V^s}({p_{t + 1}})$ and ${V^s}({p_{t}})$ denote the seed value of $p_{t+1}$ and $p_{t}$. We use a four-tuple ($p_t, a_t, p_{t+1}, r_t$) to denote a path transition.

\textbf{Expected sequence reward.} In the path transition model, the path transitions caused by the actions selected according to the policy will affect the subsequent path transition sequences and thus affect reaching the target site. To evaluate the contribution of path transitions to reaching the target site, we define the expected sequence reward as \textit{the expected sum of rewards of the path transition sequences generated by the fuzzer following a certain policy.} It can be computed recursively using the Bellman equation \cite{bellman}:
  \begin{equation}
  	{Q_\pi }(p,a) = \mathop E\limits_{p'\sim P} [r(p,a,p') + \gamma V_\pi ^t(p')]
  	\label{eq4}
  \end{equation}
  Where $ p'\sim P $ denotes the probability of transferring from path $p$ to path $p'$. $\gamma$ represents the discount factor, and the influence of subsequent path transitions on the expected sequence reward will gradually decrease. $r(p,a,p')$ denotes the reward of the path transition, ${V_\pi ^t}(p')$ denotes the transition value of path $p'$. 
 The calculation of the transition value of $p'$ is: 
  \begin{equation}
  	{V_\pi ^t}(p') = \left\{ \begin{array}{l}
  		\; \; \;\;\;\;\;\;\;\;\;\;0,\;\;\;\;\;\;\;\;\;\;if\;\;p=p_{ter}  \\
  		{\sum\nolimits_a {\pi (a|p) \cdot Q} _\pi }(p',a), \; others
  	\end{array} \right.
  	\label{eq3}
  \end{equation}
  Where $p_{ter}$ denotes the terminal path. We consider the path as the terminal path if all actions of the path can only cause self-path-transition. 
  $\pi$ represents the policy to select actions (e.g., in AFL, the fuzzer employs the random policy to select actions for mutation). $\pi (a|p)$ represents the probability of path $p$ selecting action $a$ under the policy $\pi$. If $p'$ is the terminal state, its transition value is 0. If not, its transition value is equal to the weighted average of the expected sequence rewards of all actions. 
The Bellman equation is used to recursively update the transition value and expected sequence reward until the path is the terminal path. By maximizing the expected sequence reward, the fuzzer can learn a policy that maximizes the long-term cumulative reward.

\section{METHOD}
\subsection{Overview of DeepGo}
\begin{figure*}[tp]
	\setlength{\abovecaptionskip}{0em}
	\centering
	\includegraphics[width=17.8cm,height=5.3cm]{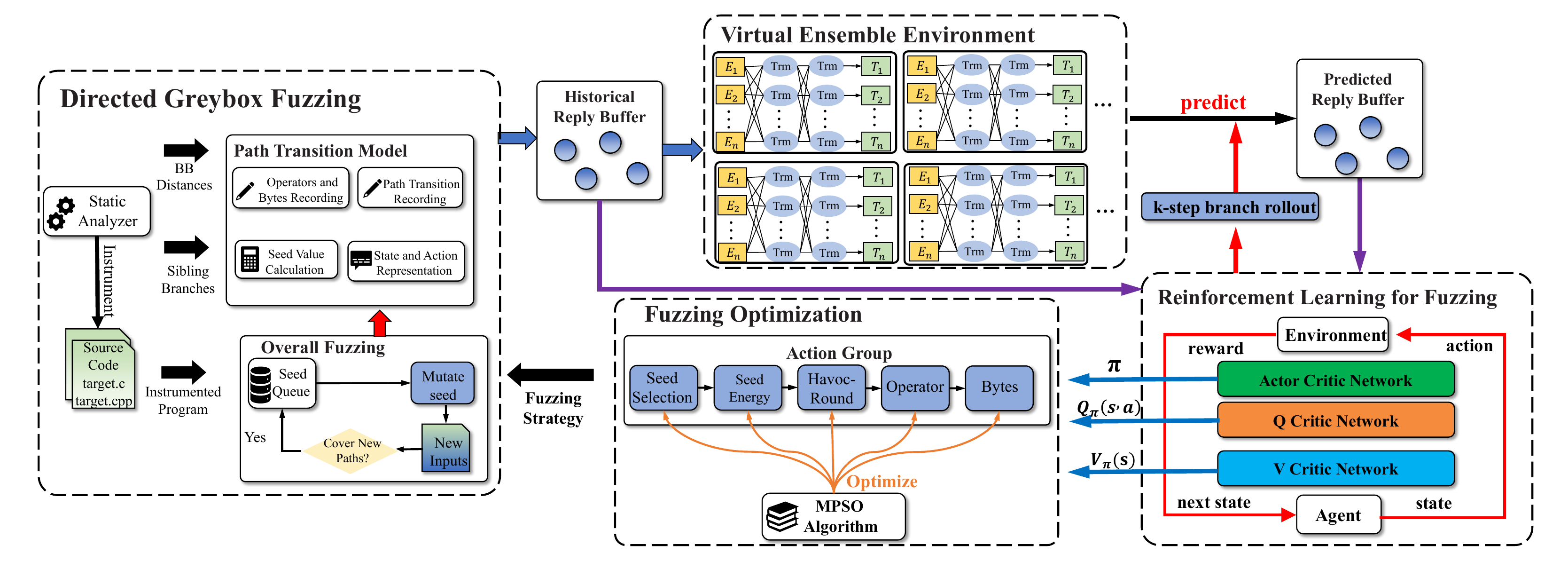}
	\caption{\label{fig:1}  The overview of DeepGo. }
	\vspace{-1.5em} 
\end{figure*}

Based on the path transition model, we design a predictable directed greybox fuzzer, DeepGo. DeepGo uses DNNs to predict potential path transitions and the corresponding rewards.
Then, it uses reinforcement learning to combine historical and predicted path transitions to obtain the optimal path transition sequence with the corresponding policy. Finally, based on the action group, it optimizes the fuzzing strategies comprehensively to exercise the optimal path transition sequences.
As Fig. \ref{fig:1} shows, DeepGo mainly consists of four components.

\textbf{Directed Greybox Fuzzing Component}. The DGF component continuously mutates seeds to generate inputs for reaching the target sites. This component contains a static analyzer and a fuzzer. At compile time, the static analyzer calculates the basic-block-level distance (BB distance), records the sibling branches of each branch, and instruments the target program. Once the fuzzing campaign is launched, the fuzzer continuously mutates seeds to test the program. Notably, the path transition model is incorporated into the DGF component.

\textbf{Virtual Ensemble Environment}. VEE is used to predict the potential path transitions and the corresponding rewards.
VEE consists of DNNs and shares the \textit{historical reply buffer} and \textit{predicted reply buffer} with the Reinforcement Learning for Fuzzing component. The historical replay buffer and predicted reply buffer are both data buffers that store the four-tuples, i.e., ($path, action, next\_path, reward$). Given a tuple, ($path, action$), the trained VEE would predict the next path and reward of the action, represented as ($next\_path, reward$), according to the probabilities of different path transitions.

\textbf{Reinforcement Learning for Fuzzing Model}. This model leverages the reinforcement learning model to combine the historical path transitions and predicted path transitions to learn the policy that maximizes sequence rewards. The RLF model consists of the \textit{Actor network}, \textit{Q-Critic network}, and \textit{V-Critic network}. 
After training, the Q-Critic network can evaluate the expected sequence reward $Q_{\pi }(p,a)$ caused by each action, the V-Critic network can evaluate the transition value $V^t_{\pi }(p)$ of each path, and the Actor network can learn the policy $\pi$ to maximize sequence rewards.

\textbf{Fuzzing Optimization Component}. This component exercises the optimal path transition sequences by optimizing the fuzzing strategies. Based on the action group, we can comprehensively optimize critical steps of fuzzing, and we use the Multi-elements Particle Swarm Optimization algorithm to optimize the elements of an action group simultaneously.

We divide the fuzzing process of DeepGo into different fuzzing cycles, with each cycle lasting approximately 20 minutes. In each fuzzing cycle, DeepGo needs to conduct four tasks, including (1) using the fuzzer to test programs and provide historical path transitions to train VEE and RLF model, (2) VEE providing predicted path transitions to train RLF model, (3) RLF providing transition values, expected sequence rewards, and policy to the FO component, and (4) the FO component using the MPSO algorithm to optimize the action group and providing the optimization strategies of fuzzing to DGF. After DeepGo completes these four tasks, it will enter the next fuzzing cycle and repeat these four tasks.

\subsection{Virtual Ensemble Environment}
\label{VEE}
To design a policy that can guide the fuzzer to optimal paths, the fuzzer has to obtain rewards for all path transitions caused by actions taken in paths. However, the fuzzer cannot take all actions in all paths within a limited time budget (e.g., 24 hours). To predict the potential path transitions and rewards caused by actions that have not yet been taken, 
we design VEE to imitate the path transition model and predict the potential path transitions and the corresponding rewards.

\subsubsection{Input and output encoding}
Before training VEE, we should encode the path, action, and reward. We design the encoding method of VEE for two purposes.  
First, VEE predicts the potential path transitions for new paths based on the recorded path transitions already taken in explored paths.
Second, to improve VEE's training efficiency, we need to use low-dimensional vectors to represent paths and actions. For example, mapping the \texttt{trace\_bits} to a 65536-dimensional vector to represent the path will significantly slow down the training of DNNs. Thus, while ensuring that different paths and actions can be clearly distinguished, we try to use low-dimensional vectors to represent paths, actions, and rewards.

\textbf{Path}. We apply the Coupled Data Embedding (CDE) \cite{CDE} algorithm to encode the path (represented by \texttt{trace\_bit}) in AFL as a 20-dimensional continuous vector and normalize the values of each dimension. CDE is used to represent discrete vectors as continuous vectors while preserving the main features that distinguish paths. According to the method of CDE, a 20-dimensional vector can both distinguish the main features of different paths and reduce the dimension of paths to improve the training efficiency of DNNs. If two paths represented by \texttt{trace\_bits} are more similar, the Euclidean distance between the corresponding 20-dimensional vectors is closer.

\textbf{Action}. We encode the action based on the mutation location. Take the example from Fig. 1, seed $s_1$ represents a seed of 100 bytes, if the fuzzer selects the 4$^{th}$ byte of $s_1$ as the mutation location, regardless of which mutator is selected, the fuzzer will mutate $s_1$ starting from the 4$^{th}$ byte, and the encoding of this action is 4/100=0.04. 
The value of all actions is also normalized.

\textbf{Reward}. The reward is a scalar which is calculated according to Equation \ref{eq5}.

Employing this encoding method enables VEE to predict the probability and reward of potential path transitions based on historical fuzzing information. VEE is trained with the four-tuples, namely ($path, action, next\_path, reward$), in which action represents the bytes of a seed where the mutations can occur, not a concrete mutation. Thus, the same action with different mutators would cause different path transitions with different probabilities. 
By analyzing the historical information of mutations on one byte, VEE can predict the path transitions caused by taking this action, i.e., mutating this byte with different mutators.
For bytes that have been mutated, we use VEE to predict the probabilities and rewards of different path transitions caused by mutating the byte using different mutators. For seeds or bytes that have not been mutated yet, we use seeds with similar structures or bytes with similar offsets to predict the probabilities and rewards of path transitions caused by mutation actions. The similarity is measured based on the CDE encoding method. If the 20-dimensional vectors representing two seeds have a shorter Euclidean distance, it indicates that the two seeds are more similar. Additionally, within a seed, similar encoded values of two actions indicate a similar offset of the corresponding bytes.

\subsubsection{Training of VEE}

We use DNN to construct VEE to imitate the path transition behavior in the path transition model. Formally, let $f$:($path, action$) $\longrightarrow$ ($next\_path, reward$) denote the DNNs that takes the tuple ($path, action$) as input and outputs the tuple ($next\_path, reward$). We use $\theta$ to denote the trainable weight parameters of DNN and train DNN with a set of training samples (X, Y), where X denotes a set of inputs and Y denotes the corresponding outputs. 
In the path transition model, since the same action on the same path may cause different path transitions with different probabilities, the path transition model is essentially a probabilistic model. Therefore, when designing DNN $f$:($path, action$) $\longrightarrow$ ($next\_path, reward$) and loss function, we mainly address VEE's aleatoric and epistemic uncertainties to improve its prediction accuracy.

\textbf{Aleatoric uncertainty}. Aleatoric uncertainty arises from the unpredictability of path transitions in fuzzing. For example, using different mutators for the same mutation location may cause different path transitions. To capture the aleatoric uncertainty, we use the Gaussian probability distribution of the next paths and rewards to predict probabilities of different path transitions and the corresponding rewards that may be caused by action $a_t$ taken in path $p_t$.  We use the trainable weight parameters to represent the Gaussian probability distribution of the next path $p_{t+1}$ and reward $r_t$ can be represented according to the input tuple ($p_t, a_t$):
\begin{equation}
	P({p_{t + 1}, r_t}|{p_t},{a_t},\theta ) = N({\mu _\theta }({p_t},{a_t}),{\Sigma _\theta }({p_t},{a_t})) 
	\label{eq7}
\end{equation}
Where $N$ represents the Gaussian distribution and ${\mu _\theta }$ represents the mean of Gaussian distribution. We use ${\Sigma _\theta }$ to represent the variance, which indicates uncertainty about the mean. We define the loss function between the output of the DNN and the ground truth label $y \in Y$ in the training set as:
\begin{equation}
    \scriptsize
	\begin{split}
	{\cal L}(\theta ) \! \! = \! \!\sum\limits_{n = 1}^N {{{[{\mu _\theta }({s_n},{a_n})\! - \!{s_{n + 1}}]}^T}\Sigma _\theta ^{ - 1}({s_n},{a_n})} [{\mu _\theta }({s_n},{a_n})\! - \!{s_{n + 1}}] \\ + \log \det {\Sigma _\theta }({s_n},{a_n}) 
	\end{split}
    \label{eqloss}
    \scriptsize
\end{equation}
The training task is to find the weight parameters $\hat{\theta}$ of the DNN $f$ to minimize the loss.

\textbf{Epistemic uncertainty}. Epistemic uncertainty results from the random sampling method employed by most DNNs. Since a single DNN cannot sample all the training data, there may be areas where the DNN has epistemic defects and cannot accurately predict outputs. To capture epistemic uncertainty, we adopted the Probabilistic Ensembles with Trajectory Sampling algorithm (PETS) \cite{PETS} to aggregate all DNNs into a virtual ensemble environment. We use the same random sampling method to train DNNs by sampling four-tuples from the historical replay buffer. All DNNs generate predicted outputs represented by Gaussian probability distributions, and their results are weighted to produce a final prediction. This prediction can be described as: 
\begin{equation}
	P({p_{t + 1}},{r_t}|{p_t},{a_t},\theta ) = {1 \over n}\sum\limits_{i=1}^{n} {P({p_{t + 1}},{r_t}|{p_t},{a_t},{\theta _i})}
	\label{eq9}
\end{equation}
Where $n$ represents the number of DNNs in VEE. Taking into account the training speed, GPU memory limit, and VEE's prediction accuracy, we use 6 identical DNNs to construct VEE and adopt the average of the probabilities from 6 DNNs as the model prediction. By taking the weighted average of all DNNs, we can alleviate the cognitive defects of a single DNN due to sampling randomness in specific areas.

\subsubsection{Determine the path transition from the predicted distribution}
Given an input ($p_t, a_t$), DNNs output the potential path transitions and the corresponding rewards, denoted as ($p_{t+1}, r_t$), with a Gaussian probability distribution.
However, in DGF, mutating the seed will only cause a specific path transition. Thus, we use a random sampling method to determine the next path $p_{t+1}$ based on the predicted probabilities of different path transitions.

 \subsection{Reinforcement Learning for Fuzzing Model}
\label{RLF}

To steer the fuzzer toward the target site more efficiently, namely through the high-reward path transition sequences, we need to learn a policy to select actions to maximize sequence rewards.
Due to the huge number of paths, actions, and path transitions in the path transition model, it is difficult to use traditional mathematical methods such as Dynamic Programming (DP) \cite{DP} to learn the policy and calculate the expected sequence rewards. Therefore, we develop the Reinforcement Learning for Fuzzing (RLF) model, which is based on the reinforcement learning algorithm Soft Actor-Critic (SAC) \cite{SAC}.

\subsubsection{The design of RLF model}
\label{designRLF}
As Fig. \ref{fig:1} shows, following the structure of SAC, the RLF model consists of a \textit{Actor network}, a \textit{Q-Critic network}, and a \textit{V-Critic network}. During the training process, we train the Q-Critic network to evaluate the expected sequence reward and train the V-Critic network to evaluate the transition value of each path. With the expected sequence rewards and transition values, the Actor network is trained to optimize the policy to increase the probability of selecting actions with high expected sequence rewards. 

\subsubsection{Training data processing and collecting}

 we treat the path transition model as the environment in reinforcement learning and map the \textit{fuzzer}, \textit{path}, \textit{action}, \textit{path transition}, and \textit{reward} in the path transition model to \textit{agent}, \textit{state}, \textit{action}, \textit{state transition}, and \textit{reward} in reinforcement learning. Moreover, RLF reuses VEE's encoding method for path, action, and reward. Since we combine historical path transitions and predicted path transitions to train RLF to give the RLF model foresight, we collect historical path transitions and predicted path transitions in different ways.
 
 \textbf{Collecting historical path transitions.} Since the environment of DGF is different from that of the traditional reinforcement learning process where state transitions are serially generated and represented as ($s_0, a_0, s_1, a_1, ..., a_{n-1}, s_n$), the training data processing of RLF is different from that of the traditional reinforcement learning (e.g., SAC). In DGF, the fuzzer might mutate one seed multiple times, resulting in different path transitions from the same path. This means that the fuzzer may stay on a certain path and take different actions to cause different path transitions. Therefore, it is not feasible for the RLF model to obtain a complete path transition sequence and evaluate the current policy's effectiveness by calculating its sequence reward within a short period (e.g., several seconds). Based on this consideration, in each fuzzing cycle,  the fuzzer selects actions according to the policy to cause path transitions. The historical path transitions are stored in the historical reply buffer and loaded by the RLF model at the end of the fuzzing cycle to train RLF during the next fuzzing cycle. 

\begin{figure}[t]
	\setlength{\abovecaptionskip}{-0.1cm}
	\centering
	\noindent \includegraphics[width=\columnwidth]{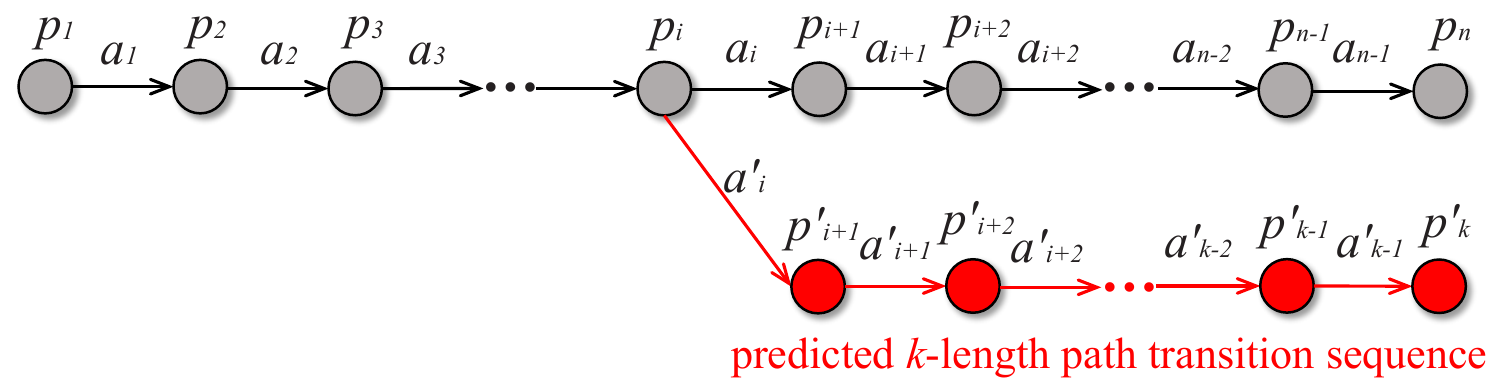}
	\caption{\label{fig:2}   k-step branch rollout strategy.}
 \vspace{-1em}
\end{figure}

 \textbf{Collecting predicted path transitions.} We use VEE to imitate the path transition model and employ the k-step branch rollout strategy to obtain predicted path transitions that have not yet occurred in the path transition model. In the k-step branch rollout strategy, the RLF model is regarded as the agent and it selects a sequence of actions at each path according to the initial policy to cause $k$ path transitions, generating a new k-length path transition sequence.
We use Fig. \ref{fig:2} to illustrate the process of the k-step branch rollout strategy. 
Suppose ($p_0, a_0, p_1, a_1 ... p_i, a_i, ... a_{n-1}, p_n$) is a historical path transition sequence in the path transition model. We take $p_i$ as the starting point and use RLF's policy $\pi$ to select a sequence of actions $a_{i}', a_{i+1}',..., a_{i+k-1}'$ to cause $k$ path transitions, generating a new $k$-length path transition sequence represented as ($p_i, a_{i}', r_{i}, p_{i+1}'$), ($p_{i+1}', a_{i+1}', r_{i+1}', p_{i+2}'$), ... , ($p_{i+k-1}', a_{i+k-1}', r_{i+k-1}', p_{i+k}'$). 
Here, $k$ is a hyperparameter that would impact the accuracy of VEE's predictions and the foresight of the RLF model's designed policy.
\graphicspath{{images/} }
 
The predictions may need to cover multiple test cases during the k-step branch rollout process. Across different test cases, the same actions defined by the locations of bytes may be \textit{misaligned} (i.e., the location of the mutated bytes in one testcase might be misaligned with the mutated bytes in other test cases owing to the different testcase length).
To handle the misalignment, we use $mutation\_ID \wedge (path\_ID>>2)$ to differentiate between different intermediate testcases generated during the process. For example, for testcase1 and testcase2 that cover the same path, we still assign them different IDs, so that we can distinguish mutations on different intermediate testcases. In this way, even if the same byte of testcase1 and testcase2 is mutated, we do not consider the probabilities and rewards associated with the path transitions caused by these two mutations to be the same.

\subsubsection{Training of the RLF model} 
The RLF model combines historical path transitions and predicted path transitions to train the Actor network, Q-Critic network, and V-Critic network. In DGF, we aim for the fuzzer to adopt actions with high expected sequence rewards while also exploring different actions to cause new path transitions. Therefore, in RLF, we apply \texttt{entropy} to measure the randomness of selecting actions. Assuming we select actions in state $s_t$ based on the policy $\pi$, and the probabilities of selecting actions follow a distribution denoted as $\pi ( \cdot |{s_t})$, the entropy of the action is calculated as:
\begin{equation}
	H(\pi ( \cdot |{s_t})) = \mathop \mathbb{E} \limits_{a \sim \pi ( \cdot |{s_t})} [ - \log\pi ( \cdot |{s_t})(a)]
	\label{eq20}
\end{equation}
In the RLF model, the objective of the Actor network is to learn a policy ${\pi ^*}$ that maximizes the reward and entropy.
\begin{equation}
	\scriptsize
		\pi^{*} \!= \! \arg \max _{\pi} \mathbb{E}_{\left(s_{t}, a_{t} \right) \sim \rho_{\pi}}[\sum_{t} \underbrace{r\left(s_{t}, a_{t}, s_{t+1}\right)}_{\text {reward }}+\alpha \underbrace{H\left(\pi\left(\cdot \mid s_{t}\right)\right)}_{\text {entropy }}]
		\label{eq30}
	\scriptsize
\end{equation}
Where $\alpha$ is the coefficient that balances exploration and exploitation, as proposed in SAC. 
Then, we construct the objective function for the Q-Critic network and the objective function for V-Critic network to train the parameters $\omega$ of the Q-Critic network and the parameters $\phi$ of the V-Critic network.
\begin{equation}
{J_{{v^\pi }}}(\phi ) = {1 \over 2}{(\gamma {r_t} + V_\phi ^\pi ({s_{t + 1}}) - V_\phi ^\pi ({s_t}))^2}
  \label{eq31}
\end{equation}
\vspace{-1em}
\begin{equation}
{J_{{Q^\pi }}}(\omega) = {({r_t} + \gamma Q_\omega ^\pi ({s_{t + 1}},{a_{t + 1}}) - Q_\omega ^\pi ({s_t},{a_t}))^2}
  \label{eq32}
  \vspace{-0.5em}
\end{equation}

For Actor network, we train the parameters $\sigma$ of the Actor network to maximize the expected sequence rewards of actions, so as to  maximize the sequence rewards. 
\begin{equation}
\vspace{-0.5em}
{J_\pi }(\sigma ) = {\max _\sigma }Q({s_t},{\pi _\sigma }( \cdot |{s_t}))
  \label{eq33}
\end{equation}
By minimizing these three objective functions, we train the parameters in the three networks to compute the expected sequence rewards, and transition values, and design the policy for selecting actions that can maximize the sequence rewards. The well-trained RLF model provides two types of optimization information to the FO component: (1) the estimated expected sequence rewards and estimated transition value, and (2) the policy for selecting actions in each path.

\subsection{Optimize Fuzzing Strategies Based on Action Group}
\label{FO}
To guide the fuzzer to exercise the optimal path transition sequences with the highest sequence rewards, we need to optimize fuzzing strategies based on the feedback information from the RLF model. In recent years, state-of-the-art techniques have been proposed to optimize a single fuzzing strategy, such as seed selection \cite{43}, mutator schedule\cite{MOPT}, and mutation location selection\cite{NEUZZ}.  
However, optimizing a single fuzzing strategy may not significantly guide the fuzzer toward the optimal path transition sequences. 
Therefore, we propose the concept of the action group that is composed of five elements and attempt to comprehensively optimize multiple fuzzing strategies. 
Besides, we propose the Multi-elements Particle Swarm Optimization (MPSO) algorithm to optimize the elements in the action group simultaneously.

\subsubsection{The concept of action group}
We define the action group as a tuple consisting of five elements.

\textbf{Seed-selection} (denoted as \textbf{SS}). Representing the probability of a seed being selected to fuzz by the fuzzer. 

\textbf{Seed-energy} (denoted as \textbf{SE}). Representing the energy assigned to the seed, which determines the number of mutations that can be applied to the seed during the havoc stage.

\textbf{Havoc-round} (denoted as \textbf{HR}). Representing the number of looping rounds used to select different mutators and bytes during the havoc stage. All of the selected mutators and bytes are integrated into a single havoc action. The value of the havoc-round may be 2, 4, 8, 16, 32, 64, or 128.

\textbf{Mutator} (denoted as \textbf{MT}). Representing the mutator selected to mutate the seed. Similar to AFL \cite{5}, DeepGo preserves 16 different types of mutators.

\textbf{Location} (denoted as \textbf{LC}). Representing the mutation location of the seed that is selected to mutate. 

\begin{figure}[t]
	\setlength{\abovecaptionskip}{0em} 
	\centering
	\noindent \includegraphics[width=\columnwidth]{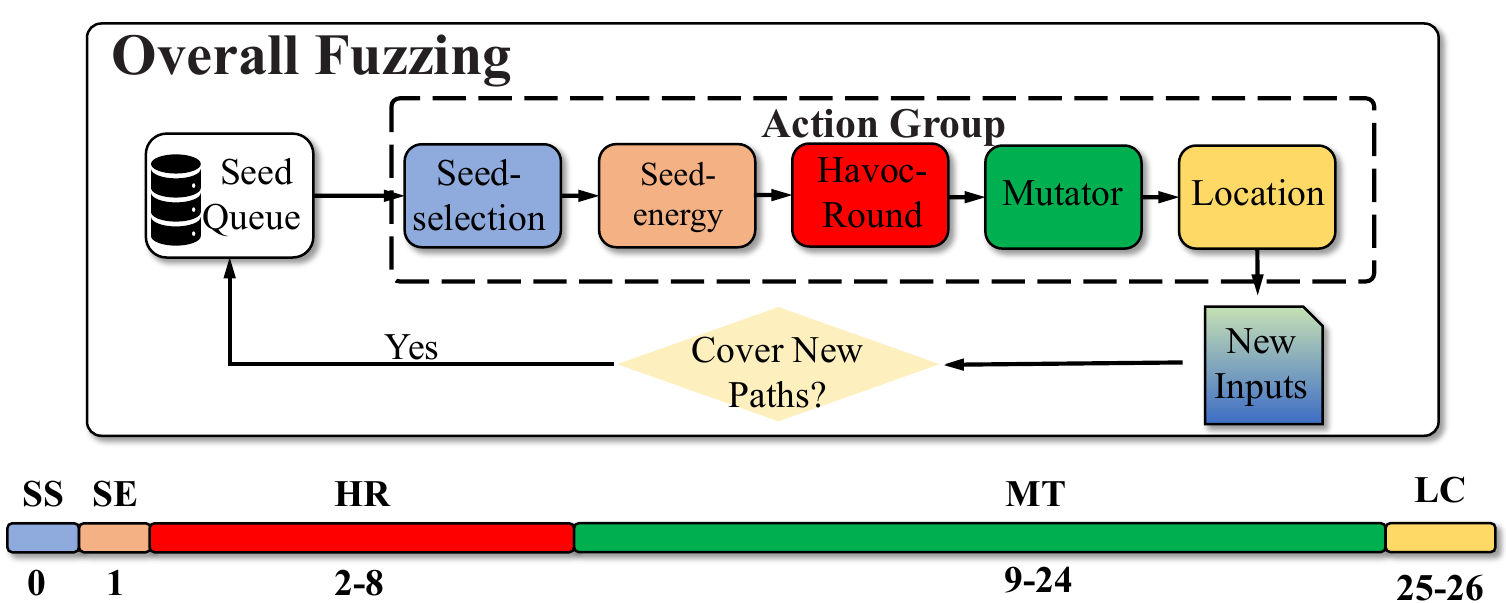}
	\caption{\label{fig:3} The layout of elements in the action group. }
	\vspace{-1.5em} 
\end{figure}

Each action group is represented as a 27-dimensional vector consisting of the 5 elements. As shown in Fig. \ref{fig:3}, SS and SE are both 1-dimensional vectors. The value of \textbf{SS} represents the probability within a range of [0, 1]. The fuzzer would select seeds to fuzz based on \textbf{SS}. The value of \textbf{SE} represents the energy assigned to the seed and the fuzzer calculates the mutation times of seeds based on \textbf{SE}. \textbf{HR} is a 7-dimensional vector, where each dimension represents the probability of selecting one of the seven different havoc-round values (i.e., 2, 4, 8, 16, 32, 64, and 128). The fuzzer samples the number of looping rounds used to select different mutators and mutation locations during the havoc stage based on \textbf{HR}. \textbf{MT} is a 16-dimensional vector, where each dimension represents the probability of selecting one of the 16 different types of mutators. \textbf{LC} is a 2-dimensional vector, where the first dimension represents the probability of selecting the optimal locations, and the second dimension represents the probability of selecting common locations. We classify the mutation locations of seeds into two categories: \textit{optimal locations} and \textit{common locations}. The optimal locations include the mutation locations selected by the RLF model's policy with a probability of greater than 80\%,  while common locations include all other mutation locations. Based on the \textbf{MT} and \textbf{LC}, the fuzzer samples the mutators and the type of mutation location. The fuzzer constantly mutates seeds to generate new inputs according to the five elements in the action group. As Fig. \ref{fig:3} shows, we represent each element as a vector and concatenate the five vectors into one vector to represent the action group of a seed.

\subsubsection{Multi-elements Particle Swarm Optimization algorithm}

To optimize the five elements in the action group simultaneously, 
we treat each action group as a particle represented by a 27-dimensional vector and view the optimization of the action group as a multi-element optimization problem. We propose a Multi-elements Particle Swarm Optimization (i.e., MPSO) algorithm to realize the optimization. The optimization of the action group can guide the fuzzer toward the desired path transition sequences with high sequence rewards.

As we have introduced in Section \ref{background}, the PSO algorithm uses massless particles to simulate birds in a flock. Each particle searches for the optimal solution individually and records it as the local best value. The local best values are then shared among the particles in the entire swarm to find the global best value as the optimal solution. These particles have two primary attributes: \textit{velocity} and \textit{position}. Velocity represents the speed of movement, and position represents the direction of movement. All particles in the swarm would constantly adjust their velocity and position based on the local best value and the global best value shared in the entire swarm: 
\begin{equation}
	{v_i} = w \times {v_i} +  r \times (lbes{t_i} - {x_i}) +  r \times (gbes{t_i} - {x_i})
	\label{eq12}
\end{equation}
\begin{equation}
	{x_i} = {x_i} + {v_i}
	\label{eq13}
\end{equation}
Where ${v_i}$ represents the velocity of the $i^{th}$ particle, ${x_i}$ represents the position of the $i^{th}$ particle, $r$ represents the random displacement weight within the range of [0,1], $lbest_i$ represents the local best position found by the $i^{th}$ particle and $gbest_i$ represents the global best position found by all particles. The inertia factor $\omega$ is a non-negative value, where a larger value results in stronger global optimization and weaker local optimization ability in the PSO algorithm.  In this paper, we apply the Linearly Decreasing Inertia Weight (LDIW) method \cite{LDIW} to set the value of $\omega$ as:
\begin{equation}
	\omega  = {\omega _{ini}} - ({\omega _{ini}} - {\omega _{end}}) \times (g/G)
	\label{eq14}
\end{equation}
Where $\omega_{ini}$ and $\omega_{end}$ denote the initial and final inertia values, respectively. Following the LDIW, $\omega_{ini}$ and $\omega_{end}$ are typically set to 0.4 and 0.9. $g$ denotes the number of iterations completed by the current particle, which is equal to the number of mutations performed by the fuzzer on the corresponding seed.
$G$ denotes the maximum iteration number, which is equal to the total mutation times calculated according to the seed energy. In DeepGo, the mutation times of seeds are determined by the energy assigned to the seeds.

\textbf{Local best position and local efficiency}. In the MPSO algorithm, each particle has its own local best position \texttt{lbest} and local efficiency $eff_{local}$. Given a particle, position $x_1$ is deemed superior to position $x_2$ only if the local efficiency obtained by the particle at $x_1$ is greater than that at $x_2$. We use local efficiency to quantify the mutation efficiency of fuzzing strategies composed of the five elements for one specific particle.
The local efficiency is calculated according to the averaged expected sequence rewards of all the mutations taken by the fuzzer in the path corresponding to the particle. 
In DGF, each mutation on the seed would cause a path transition from path $p$ to path $p'$. According to Equation \ref{eq4}, in one path transition, we use ${r+\gamma{V^t_\pi}({p'})}$ to evaluate the expected sequence rewards. Based on this, we calculate local efficiency.
\begin{equation}
\vspace{-0.5em}
	ef{f_{local}} = {{\sum\limits_{i = 1}^g {r_i+{\gamma V^t_\pi}({p'})} } \over g}
	\label{eq17}
 \vspace{0.5em}
\end{equation}
Where ${V^t_\pi}({p'})$ denotes transition value of path $p'$ covered by the $i^{th}$ mutation.

\textbf{Global best position and global efficiency}. We use global efficiency to quantify the fuzzing efficiency of the fuzzer. Since the global efficiency of the fuzzer depends on the local efficiency of different particles and the relationship between particles, we determine the global position of all particles based on the fuzzing efficiency within a fuzzing cycle. A particle is currently in the global best position (\texttt{gbest}) only if the fuzzing efficiency of the fuzzer is higher than that of any other position.
We use the average sequence rewards during the fuzzing cycle to evaluate the global efficiency:
\begin{equation}
\vspace{-1em}
	ef{f_{global}} = {{\sum\limits_{j = 1}^U {\sum\limits_{i = 1}^{{g_j}} {r_i+{\gamma V^t_\pi}({p'_j})} } } \over {\sum\limits_{j = 1}^U {{g_j}} }}
	\label{eq18}
\vspace{0.5em}
\end{equation}
Where  $eff_{global}$ denotes the global efficiency, $p_j$ denotes $j^{th}$ seed, $g_j$ denotes the total number of mutations for $p_j$, and $U$ denotes the number of seeds that have been fuzzed in the current fuzzing cycle. 

During the fuzzing process, we calculate the local efficiency and global efficiency of particles according to Equations (\ref{eq17}) and (\ref{eq18}), and record the \texttt{lbest} and \texttt{gbest}. We update the particles' spatial positions according to  Equations (\ref{eq12}), (\ref{eq13}), and (\ref{eq14}), moving all particles towards the direction of \texttt{lbest} and \texttt{gbest}. Through this approach, we optimize the action groups of all seeds to guide the fuzzer toward the target via the optimal path transition sequences. 

\begin{algorithm}[t]
	\footnotesize
	\caption{MPSO Algorithm}
	\label{alg:MPSO}
	\begin{algorithmic}[1]
		\REQUIRE $\Omega_{(s, p)}$
		\ENSURE $U_s$,  $\Omega_{(s, p')}$
		\STATE Initial($\Omega_{(s, p)}$)
		\WHILE {$fuzzing$}
		  \FOR{$(s_i, p_i) $ in  $\Omega_{(s, p)}$}
		    \IF {$Prob\_Sel_s$($p_i$(\textbf{SS})) == \textbf{True}}
		       \STATE \textit{$mn_i$} $\leftarrow $\texttt{Cal\_MN}($p_i$(\textbf{SE}))
		       \FOR{$j $ in  $mn_i$}
		          \STATE \textit{$hr_j$} $\leftarrow \$Prob\_Sel_h$($p_i$(\textbf{HR}), $hm_j  \gets <>  $ 
		          \FOR{$k$ in  \textit{$hr_j$}}
		            \STATE $lc_k$ $\leftarrow$ $Prob\_Sel_l$($p_i$(\textbf{LC})),
                        \STATE $mt_k$ $\leftarrow Prob\_Sel_m$($p_i$(\textbf{MT})), 
		            \STATE $hm_j \gets hm_j \cup {(lc_k, mt_k)}$
                  \ENDFOR
                 \STATE $new\_input$ = \texttt{Mutate}($hm_j$, $s_i$)
                 \STATE $eff_{local}, eff_{global}$ = \texttt{Cal\_eff}($s_i$, $new\_input$)
                 \STATE \texttt{Update}(\texttt{lbest}, \texttt{gbest}, \texttt{$p_i$})
               \ENDFOR
		    \ENDIF
		  \ENDFOR
		\ENDWHILE
	\end{algorithmic}
\end{algorithm}
The process of MPSO is shown in Algorithm \ref{alg:MPSO}, where $s$ denotes the seed, $p$ denotes the particle, $\Omega_{(s, p)}$ denotes the set containing all seeds and their corresponding particles. The function $Prob\_Sel_s$ determines whether to fuzz the seed based on the \textbf{SS}. $mn_i$ denotes the number of mutation times of the seed, and the function \texttt{Cal\_MN} calculates the number of mutation times for the seed based on \textbf{SE}. The functions $Prob\_Sel_h$, $Prob\_Sel_l$, and $Prob\_Sel_m$ probabilistically select the corresponding values of $hr$ for \textbf{HR}, $lc$ for mutation location based on \textbf{LC}, and $mt$ for mutator based on \textbf{MT}. 
 
At first, we initialize the five elements in all the particles (Line 1). For \textbf{SS} and \textbf{SE}, according to the seeds' characteristics (e.g., the bitmap size and execution speed), AFL \cite{5} calculates the probabilities of seeds being fuzzed and the energy assigned to seeds. We use the probabilities and the seed energy obtained by AFL's method as the initial value of \textbf{SS} and \textbf{SE}. For \textbf{HR}, \textbf{LC}, and \textbf{MT}, we use the average probability as the initial value of their spatial position (e.g., 1/16 for each mutator). 
During the fuzzing process, the fuzzer selects seeds to fuzz according to \textbf{SS} (Line 4), calculates the mutation times according to \textbf{SE} (Line 5), selects the values of \textbf{HR} (i.e., $hr_j$)  (Line 7), selects the seed bytes for mutation (i.e., $lc_k$) according to \textbf{LC} (Line 9), and selects the mutators (i.e., $mt_k$) according to \textbf{MT} (Line 10). In each $hr_j$ cycle, the $mt_k$ and $lc_k$ will be combined into the havoc mutation (i.e., $hm_j$) (Line 11), and the fuzzer will use $hm_j$ to mutate the seed $s_i$ and generate a $new\_input$ (Line 13). Then, MPSO would calculate the local efficiency and global efficiency (Line 14), and update \texttt{lbest}, \texttt{gbest} and the position of the particle $p_i$ (Line 15). Notably, the value of all dimensions of the particle will constantly change according to Equation (\ref{eq12}), (\ref{eq13}), and (\ref{eq14}) to update the spatial position of the particle, allowing the particle to move to the \texttt{lbest} and \texttt{gbest}. For instance, if one particle has low local efficiency which results in reducing the global efficiency of the fuzzer, according to the Equation (\ref{eq12}), (\ref{eq13}) and (\ref{eq14}), the \textbf{SS} and the \textbf{SE} will gradually decrease during the process of MPSO. Once the FO component first receives the feedback information from the RLF model, it will launch the MPSO algorithm until the end of fuzzing.

\section{Implementation}
The implementation of DeepGo mainly consists of three components: the fuzzer, the VEE, and the RLF model. For the static analyzer in the fuzzer, we leverage LLVM 11.0. We use the LLVM IR to instrument the program and obtain information about basic-block-level distance, sibling branches, etc. The fuzzer is built on AFLGo, with 2100 lines of C code, and the VEE and the RLF model are implemented with about 1300 lines of Python code. 

In detail, the DNNs of VEE are implemented in Pytorch-1.13.0 with five fully connected layers. The hidden layer uses Swish as its activation function. The DNNs are trained for 500 epochs and we use the tensorboard tool to automatically monitor the loss values to determine if they converge to a small value. If the loss values of VEE and RLF have converged, the DNNs will automatically stop training. The networks in the RLF model consist of three fully-connected layers. For the hyperparameters of the RLF model, referring to the learning rate settings for Q-Critic network, V-Critic network, and Actor network in SAC, we set them to 0.005 to ensure the learning efficiency and convergence of the RLF model (in the experimental process, all three networks can converge quickly).

Notably, when using DeepGo to test programs, the fuzzing process, the training of the models, and the prediction of the path transition are performed concurrently. We use an extra GPU to train the VEE and RLF models based on the information collected throughout the fuzzing process. All the time spent on training and prediction of the VEE and RLF models is counted in the time budget (indicated as wall clock time) for the fuzzing process.

\section{Evaluation}

To evaluate the effectiveness of DeepGo, we conducted experiments aiming to answer five research questions:

\noindent \textbf{RQ1:} What about the performance of DeepGo in terms of reaching the target code locations?

\noindent \textbf{RQ2:} What about the performance of DeepGo in terms of exposing the known vulnerabilities?


\noindent \textbf{RQ3:} What about VEE's performance in predicting probabilities and rewards of path transitions?

\noindent \textbf{RQ4:} Can the RLF model and FO component guide the fuzzer to path transition sequences with high sequence rewards?

\noindent \textbf{RQ5:} How do the VEE, RLF, and FO components contribute to the overall performance of DeepGo?

\subsection{Evaluation Setup}

\textbf{Evaluation Criteria.} We use two types of criteria to evaluate the performance of different fuzzing techniques.

(1) Time-to-Reach (TTR) is used to evaluate the time spent on generating the first input that can reach the target site.  

(2) Time-to-Expose (TTE) is used to evaluate the time spent on exposing the (known or undisclosed) vulnerabilities in the target sites. An observed crash indicates that the fuzzer has successfully exposed the vulnerability.

\textbf{Evaluation Benchmarks.} We selected two datasets that are widely used by state-of-the-art DGF techniques (e.g., WindRanger \cite{23}, BEACON \cite{26}).

(1) UniBench \cite{42} provides real-world programs of different types and the corresponding seed corpus. The state-of-the-art fuzzing techniques, such as WindRanger, have used the UniBench as the benchmark for testing. To answer RQ1, RQ3, RQ4, and RQ5, we tested the 20 programs from UniBench and used AFL++ \cite{AFL++} to select target sites from each program by conducting preliminary experiments. We first ran AFL++ for 48 hours and collected all the seeds generated by AFL++. Then, we use afl-cov to re-run these seeds, so that we can obtain the code locations covered and the time when they are covered, represented as pairs like (line, time). Finally, among the locations that are reached using from 1 hour to 48 hours (i.e., more than 1 hour), we randomly selected 4 code locations as the targets.

(2) AFLGo testsuite \cite{43} was proposed in AFLGo's paper and website to evaluate the directness of DGF. It had been used as a benchmark by the state-of-the-art DGF techniques to verify the bug reproduction capabilities. To answer RQ2, we selected the AFLGo testsuite as the benchmark to verify the capability of exposing known vulnerabilities.

\textbf{Baselines.} In our evaluation, we compared DeepGo with the state-of-the-art directed greybox fuzzers that were publicly available at the time of writing this paper, including WindRanger, BEACON, ParmeSan, and AFLGo.

\textbf{Experiment Settings.} We conducted the experiments on the machine equipped with Intel(R) Xeon(R) Gold 6133 CPU @ 2.50GHz with 80 cores and used Ubuntu 20.04 LTS as the operating system. All the experiments were repeated \textbf{5} times within a time budget of \textbf{24 hours}. When testing the programs from UniBench and AFLGo testsuite, we used the seeds in BenchMark's recommended seed corpus as initial seeds. 
For experimental results analysis, we utilize the Mann-Whitney U test (p-value) to measure the statistical significance. In addition, we use the Vargha-Delaney statistic ($\hat{A}_{12}$) \cite{39} to measure the probability of one technique performing better than another.

\subsection{Reaching Target Sites} 
\label{reach}
To answer RQ1, we tested the 20 programs from UniBench, with a total of 80 target sites, and evaluated the TTR of different fuzzers. We set the timeout threshold as 24 hours.  
The detailed results of TTR are listed in Table \ref{table:unibench} in the Appendix.
In Table \ref{table:unibench}, the entry ``N/A" indicates that the fuzzer failed to compile the program due to code issues, while ``T.O." indicates that the fuzzer couldn't reach the target site within the allocated 24-hour time budget. For WindRanger, some entries are marked as ``N/A" due to encountering segmentation fault errors or being unable to obtain distance information during program testing. As for BEACON and ParmeSan, most entries showing ``N/A" might be because it is incompatible with UniBench. 
For ``N/A" entries, we did not use them to calculate the speedups and p-values. 
As for the ``T.O." entries, we believe that these fuzzers might still reach the targets in subsequent fuzzing processes. Therefore, we opted for a slightly larger value of 1500 minutes to calculate speedups and p-values.

According to the results of TTR, DeepGo can reach the most (73/80) target sites compared to AFLGo (22/80), BEACON (11/80), WindRanger (19/80), and ParmeSan (9/80) within the time budget. Moreover, on most of the target sites (67/80), DeepGo outperforms all other fuzzers and achieves the shortest TTR. In terms of mean TTR of reaching the target sites, DeepGo demonstrates 3.23$\times$, 1.72$\times$, 1.81$\times$, and 4.83$\times$ speedup compared to AFLGo, BEACON, WindRanger, and ParmeSan, respectively. 
We conducted both the Mann-Whitney U test (p-value) and the Vargha-Delaney test ($\hat{A}_{12}$) that all the p-values are less than 0.05, and the mean $\hat{A}_{12}$ against AFLGo, BEACON, WindRanger, and ParmeSan are 0.86, 0.81, 0.83, and 0.89, respectively. Based on the above analysis, we can conclude that \textbf{DeepGo can reach the target sites faster than baseline fuzzers}.

\begin{figure}[t]
	\centering
	\setlength{\abovecaptionskip}{-0.1cm}
	\includegraphics[width=\columnwidth]{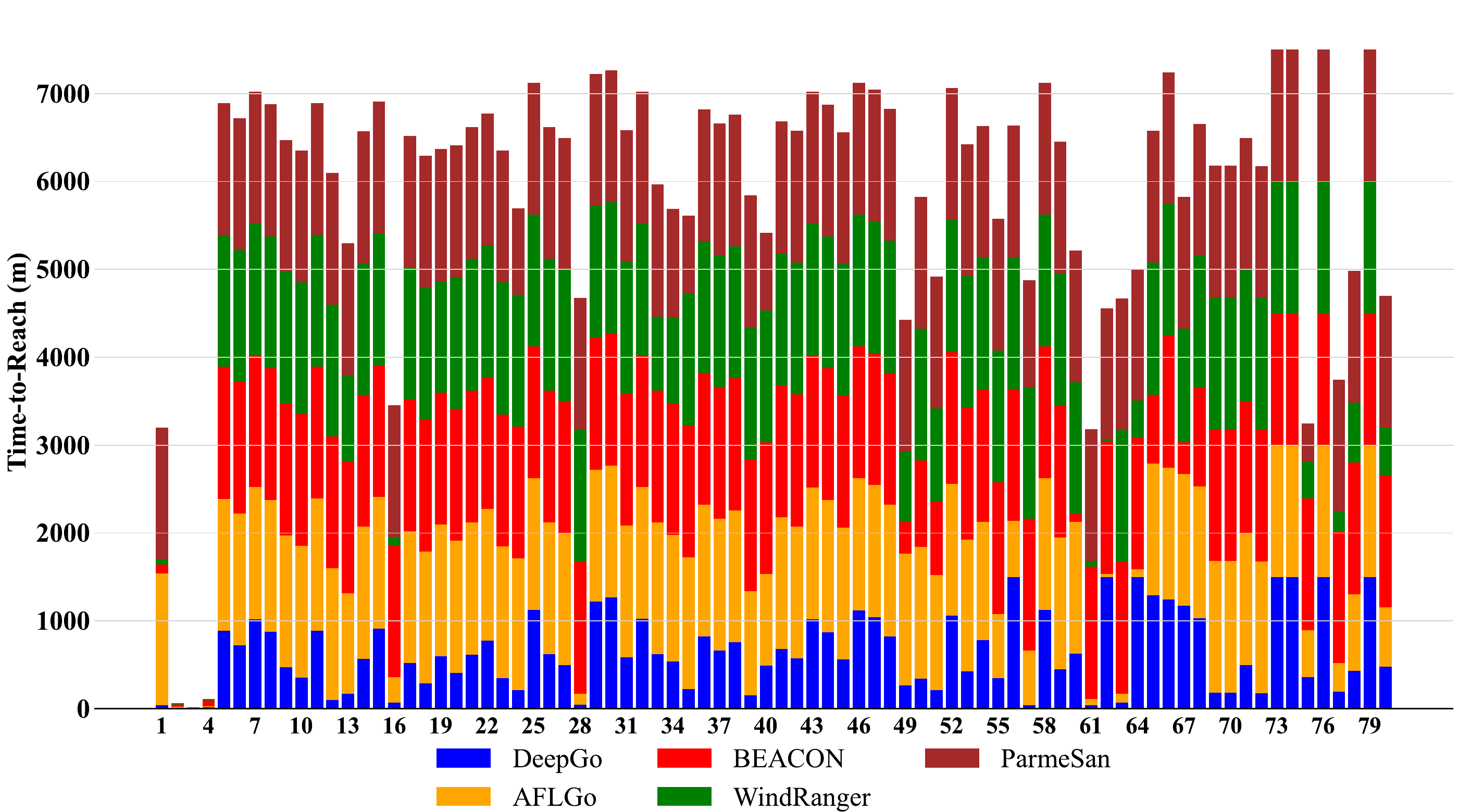} 
	\caption{\label{fig:ttr}  TTR of DeepGo and baseline fuzzers on the UniBench.} 
 \vspace{-1em}
\end{figure}

To reflect the results in a straight way, we use bar charts to visualize the results. In Fig. \ref{fig:ttr}, the x-axis represents the target site ID (1-80), the y-axis represents the total TTR of all fuzzers in minutes, 
and a shorter bar indicates a shorter TTR. Since some fuzzers cannot compile some programs or reach the target sites within the 24-hour time budget, resulting no TTR. To distinguish these cases, the TTR of such a case is represented as 1500 min in Fig. \ref{fig:ttr}. From the figure, we can clearly see that the blue bars are much shorter than the other bars, which means that DeepGo can reach most of the target sites faster than the baseline fuzzers. 

\subsection{Exposing vulnerabilities} 
\label{expose}
To answer RQ2, following BEACON and WindRanger, we used the AFLGo testsuite and set the known vulnerabilities with CVE IDs in the programs as the target sites. The information on target sites and the TTE results
are presented in Table \ref{table:tte}. As Table \ref{table:tte} shows, among the 20 vulnerabilities, DeepGo (19) exposed the most compared to AFLGo (14), BEACON (13), 
	WindRanger (16), and ParmeSan (14). Besides, on most of the target sites (14/20), DeepGo outperformed all the baseline fuzzers and achieved the shortest TTE. With respect to the mean TTE of exposing vulnerabilities, DeepGo demonstrated 2.61$\times$, 3.32$\times$, 2.43$\times$ and 2.53$\times$ speedup compared to AFLGo, BEACON, WindRanger, and ParmeSan, respectively. All p-values were less than 0.05, and the mean $\hat{A}{12}$ against AFLGo, BEACON, WindRanger, and ParmeSan were 0.79, 0.72, 0.75, and 0.81, respectively. Based on the above analysis, we can conclude that \textbf{DeepGo can expose known vulnerabilities faster than the baseline fuzzers}.
 	\begin{table}[t]

		\footnotesize
		\centering
		\setlength{\abovecaptionskip}{0.05cm} 
		\setlength{\belowcaptionskip}{0.05cm}
		\setlength{\tabcolsep}{2pt}
		\caption{The results of TTE on AFLGo testsuite}
		\label{table:6}
		
		\begin{tabular}{ccccccccc}
			\toprule[1.5pt]
			\textbf{Prog.}&   \textbf{CVE-ID}&{AFLGo}  & {BEACON} & {WindRa} & {ParmeS} & {DeepGo} \\
			\midrule
			\multirow{7}{*}{binutils$_{2.26}$}   & 2016-4487 &2.33m  & 0.63m & 1.21m & 0.95m  & 1.34m \\
			&2016-4488      &4.23m   & 32.1m & 3.32m &  2.62m  & 2.69m \\
			& 2016-4489    &3.36m & 2.98m & 5.88m  & 2.31m   & 1.23m \\
			& 2016-4490   &1.15m   & 2.35m & 2.63m  & 0.82m   & 1.97m  \\
			& 2016-4491    &448m    &  258m & 298m  &  212m  & 129m  \\
			& 2016-4492      &10.8m  & 43.6m  & 7.47m  &  4.33m  & 6.94m \\
			&   2016-6131     &348m  & 292m &  318m & 244m  & 68.1m \\ \midrule
			\multirow{4}{*}{libming$_{4.48}$} & 2018-8807 &331m  & 267m  & 171m & 301m   & 101m \\
			&2018-8962    &234m  & 163m & 121m & 198m   & 54.8m \\    
			& 2018-11095     &T.O. & 914m  & 1311m & T.O. & 812m \\
			& 2018-11225     &T.O.  & 438m & 996m  & T.O.  & 128m  \\ \midrule
			\multirow{3}{*}{LibPNG$_{1.5.1}$}  &2011-2501 &10.2m  & N/A & 7.81m & 4.53m   & 3.46m \\
			&2011-3328     &69.1m  & N/A & 49.3m & 193m   &  17.5m \\
			& 2015-8540  &0.88m  & N/A & 0.96m  & 3.41m   & 5.65m \\ \midrule
			\multirow{4}{*}{xmllint$_{2.9.4}$}  & 2017-9047 &T.O.   & T.O. & T.O. & T.O. & 783m \\
			& 2017-9048    &T.O.  & T.O. & T.O. &  T.O.  & 1389m  \\
			& 2017-9049    &T.O.  & T.O. & T.O. &  T.O.  & T.O. \\
			& 2017-9050    &T.O.  & T.O. & T.O. & T.O.   &911m \\ \midrule
			\multirow{2}{*}{Lrzip$_{0.631}$}  & 2017-8846 &348m  & 156m&  223m &  466m  & 131m \\
			&2018-11496    & 201m  & 98.1m & 169m & 126m   & 78.9m \\
			
			\midrule
			\multicolumn{2}{c}{\fontsize{6}{5}\selectfont \textbf{speedup}}         &{\textbf{2.61$\times$}}  & {\textbf{3.32$\times$}} & {\textbf{2.43$\times$}} & {\textbf{2.53$\times$}} & {\textbf{-}} \\ 
			\multicolumn{2}{c}{\fontsize{6}{5}\selectfont \textbf{mean $\hat{A}_{12}$}}      &{\textbf{0.79}}  & {\textbf{0.72}} & {\textbf{0.75}} & {\textbf{0.81}} & {\textbf{-}} \\
			\multicolumn{2}{c}{\fontsize{6}{5}\selectfont \textbf{mean p-values}} & {\textbf{$0.018$}}  & {\textbf{$0.032$}} & {\textbf{$0.026$}} & {\textbf{$0.011$}} & {\textbf{-}} \\  
			\bottomrule[1.5pt]
			
		\end{tabular}
		\label{table:tte}
  \vspace{-0.5em}
\end{table}



\subsection{The effectiveness of VEE}
\label{VEEpredict}

To answer RQ3, we analyzed the predictions made by VEE when DeepGo tested the 20 programs from UniBench. Given an input ($p_t, a_t$), VEE would predict the outputs ($p_{t+1}, r_t$) with varying probabilities and rewards (i.e., predicted probabilities and predicted rewards). Meanwhile, during the fuzzing process, the fuzzer takes action $a_t$ in path $p_t$ would cause path transitions with varying probabilities and rewards (i.e., real probabilities and real rewards.). To analyze the prediction accuracy of VEE, 
we defined the metric \textbf{Accuracy} as:
\begin{equation}
Accuracy = 1 - {{|real\_value - predicted\_value|} \over {|real\_value|}}
\label{eq20}
\end{equation}
Where $real\_value$ denotes the real probability or real reward, $predicted\_value$ denotes the predicted probability or predicted reward. The higher value of Accuracy indicates the higher accuracy of the prediction.
\begin{figure}[t]
\centering
\subfigcapskip=-2pt
\setlength{\abovecaptionskip}{-0.1cm}
\subfigure[AAPP]{
\includegraphics[width=\columnwidth]{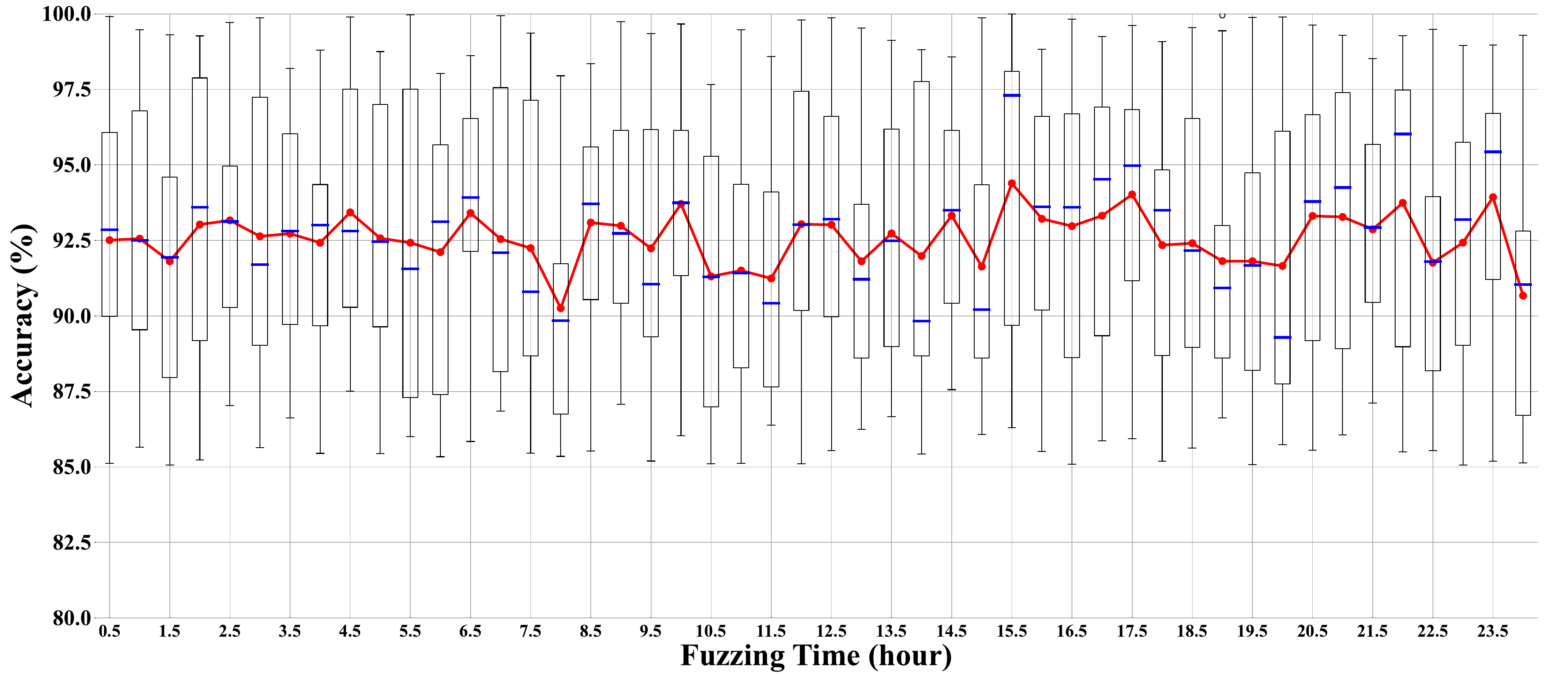}
}%
\vspace{-0.1cm}
\subfigure[AAPR]{
\includegraphics[width=\columnwidth]{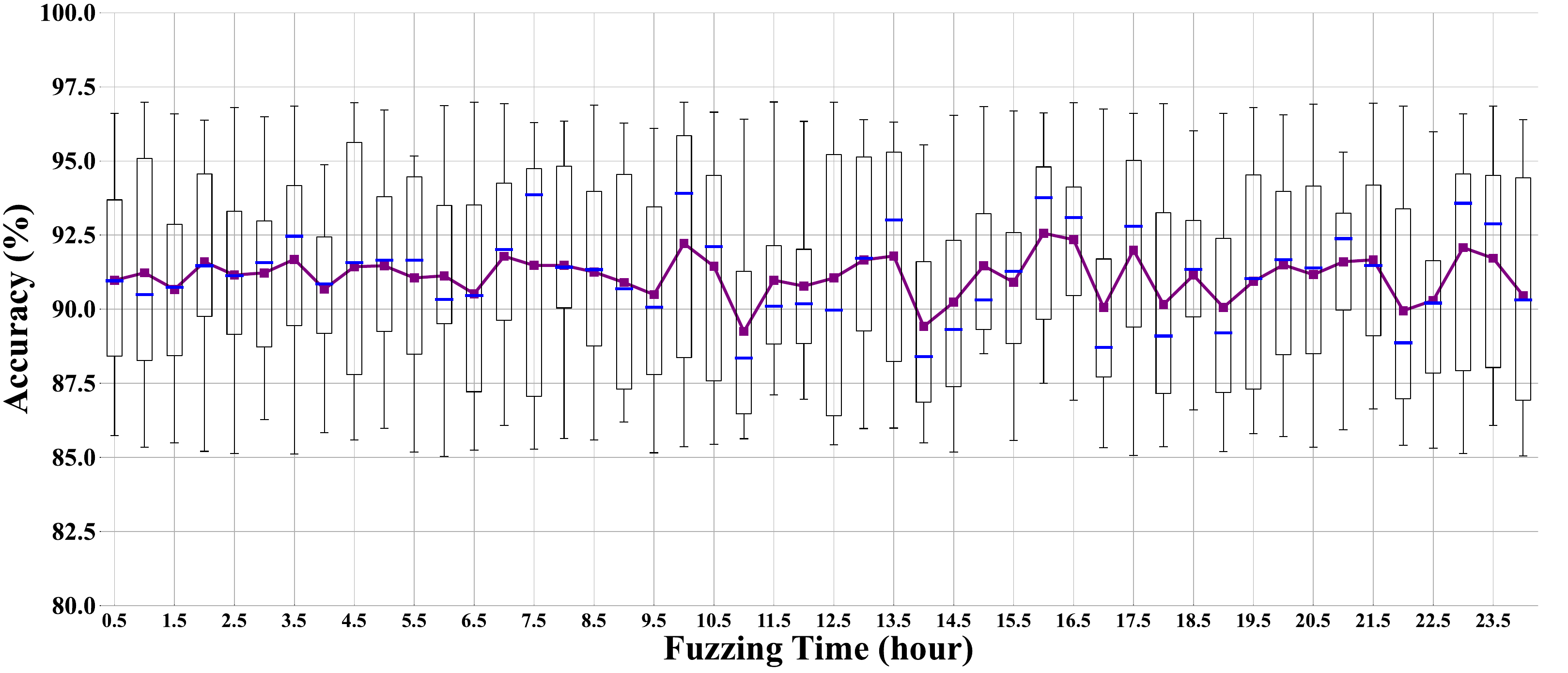}
}      
\caption{ \label{fig:VEE}  The accuracies of rewards and probabilities predicted by VEE on programs from UniBench.} 
\vspace{-1em}
\end{figure}
During the testing of each program from UniBench, we calculated the two accuracies of each path transition every 30 minutes. Since DeepGo repeated the testing of the 20 programs five times, at each time point, 
we calculated the average accuracy of predicted probability (\textbf{AAPP}) and the average accuracy of predicted reward (\textbf{AAPR}) of all path transitions of all the programs, to evaluate the prediction accuracy of VEE based on Equation \ref{eq20}. 

To reflect the results straightforward, we use the line chart to visualize the results. In Fig. \ref{fig:VEE}, the x-axis represents the fuzzing time (from 0-hour to 24-hour) and the y-axis represents the accuracies of AAPP and AAPR which are both within the range of [0, 1].  As shown in Fig. \ref{fig:VEE}, from 0.5-hour to 24-hour, 
the accuracies of AAPP and AAPR are all greater than 80\%, and the means of AAPP and AAPR at the 48 time points are 
92.57\% and 91.10\% respectively. 
Meanwhile, we used box charts to show the AAPPs and AAPRs on different programs at different time points. The blue line represents the median, and the length of the box represents the distribution range. From the box charts, we can see that the AAPPs and AAPRs do not vary significantly ($ \le  15\%$) among different programs, and maintain high value ($\ge 80\%$) at each time point.
This suggests that VEE can limit the deviation when imitating the path transition model and predict the probabilities and rewards of path transitions with high accuracy.

\begin{figure}[b]
\centering
\setlength{\abovecaptionskip}{-0.1cm}
\includegraphics[width=\columnwidth]{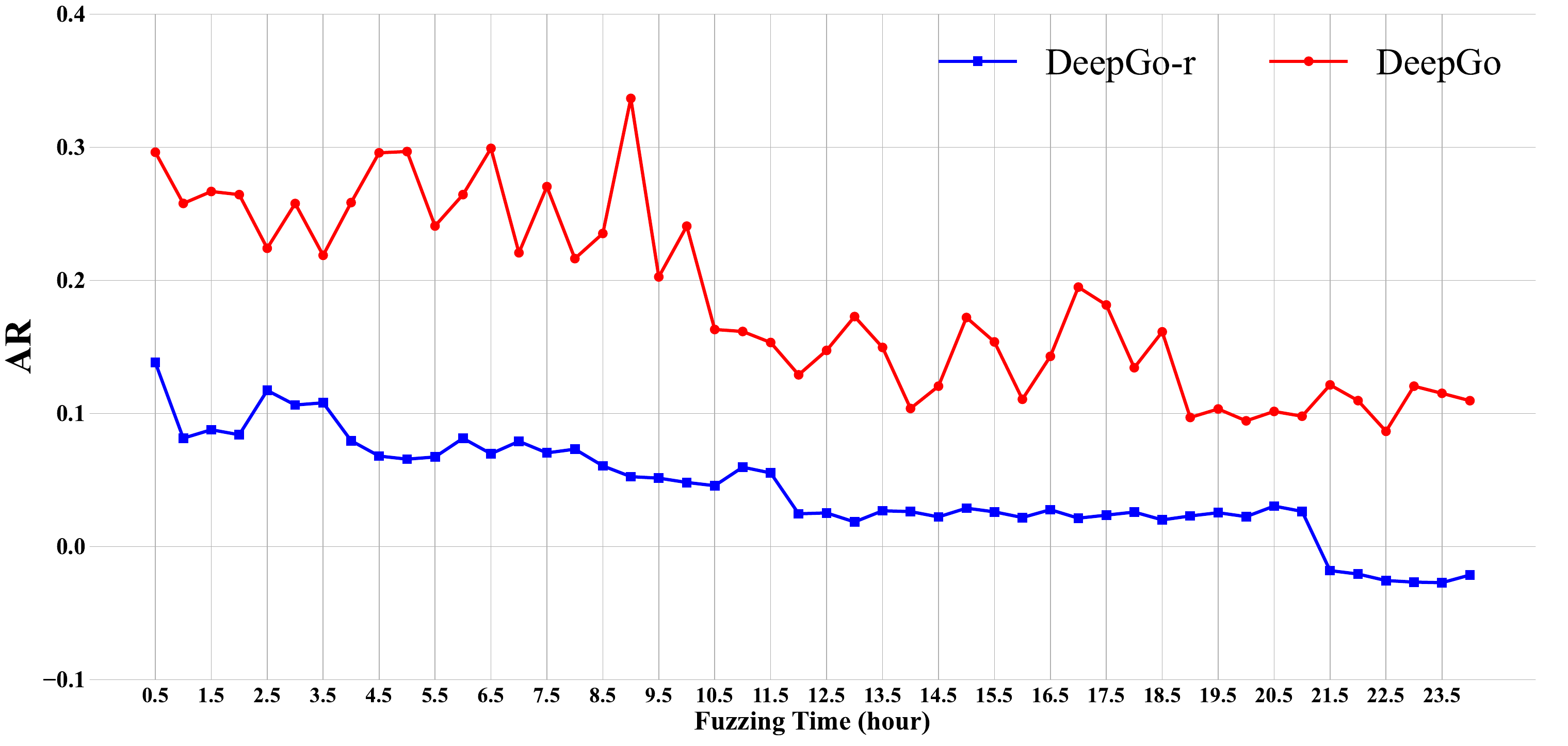} 
\caption{\label{fig:RLF}  The value of ARs on programs from UniBench.} 
\vspace{-1em}
\end{figure}

\subsection{The effectiveness of the RLF model and FO component}
\label{reward}
To answer RQ4, we collected rewards of path transitions that occurred during the fuzzing process, 
to calculate the average reward of path transitions in the path transition sequence to show whether the RLF model and the FO components can guide the fuzzer to exercise the optimal path transition sequences with high sequence rewards. 
Since DeepGo's optimizations on the fuzzing strategies rely on the RLF model and the FO component, removing either of them will render DeepGo's optimizations ineffective, thus, we removed both of them, thereby forming a new fuzzer called \textbf{DeepGo-r}.
We used DeepGo-r and DeepGo to test the 20 programs from UniBench and calculated the reward of all path transitions for all the programs every 30 minutes. Then, at each time point, we obtain the average reward (i.e., \textbf{AR}) of the 20 programs.


To reflect the results in a straight way, we use the line chart to visualize the results. In Fig. \ref{fig:RLF}, the x-axis represents the fuzzing time (from 0.5-hour to 24-hour) and the y-axis represents the values of AR, which is within the range of [-1, -1]. As shown in Fig. \ref{fig:RLF}, from 0.5-hour to 24-hour, the overall trend of AR is decreasing, which is consistent with the fact that the fuzzer is getting difficult to find new seeds as the fuzzing proceeds.
DeepGo's AR was significantly higher than that of DeepGo-r. At each time point, DeepGo's AR was, on average, 4.26$\times$ higher than DeepGo-r's AR. This indicates that the RLF model and the FO component can guide the fuzzer to exercise the optimal path transition sequences and reach the target site more quickly.

\subsection{Ablation study}
To answer RQ5, we conducted the ablation study to demonstrate the impact of VEE, RLF, and FO on DeepGo's performance.
To demonstrate that VEE, the RLF model, and the FO component can enhance directedness, we remove VEE from DeepGo and form a new tool \textbf{DeepGo-v} and also run DeepGo-v and DeepGo-r on UniBench for the TTR experiment. 
Detailed results of DeepGo-v and DeepGo-r are listed in Table \ref{table:unibench} in the Appendix. According to the TTR results, DeepGo (73/80) can reach much more target sites than DeepGo-v (32/80) and DeepGo-r (18/80), respectively. Moreover, DeepGo outperforms DeepGo-v and DeepGo-r by 2.05$\times$ and 3.72$\times$, respectively, in the average TTR of reaching the target sites. The p-values are 0.013 and 0.006, and the mean $\hat{A}_{12}$ against DeepGo-v and DeepGo-r is 0.83 and 0.90, respectively. These results demonstrate that VEE, RLF, and the FO component have significant impacts on reducing TTR.




We also use intermediate data to prove that DeepGo can avoid the infeasible and hard-to-execute paths. Firstly, we consider 
a seed to be a reachable seed (i.e., \textbf{Rseed})
if its execution path covers reachable basic blocks (BB distance $ \ge $ 0). We calculate the average number of reachable seeds generated by different fuzzers when testing different programs, as well as the proportion of reachable seeds (i.e., \textbf{PRseed}) among all
seeds. If the number and proportion of reachable seeds are higher, it indicates that the fuzzer can avoid spending time on infeasible and hard-to-execute paths. 
Secondly, we measure the total number of paths taken by the fuzzer to reach target sites (i.e., reached paths, \textbf{Rpath}) during testing 20 programs from UniBench. For the same target site, different paths can yield different testing results. 
By counting the total number of reached paths, we can determine if DeepGo can discover more paths to reach the target site compared to other fuzzers, thus enabling more diverse testing of the target site.
\begin{table}[t]
	\footnotesize
	\centering
        \setlength{\abovecaptionskip}{0.1cm} 
	\setlength{\belowcaptionskip}{0.1cm}
	\caption{Intermediate data analysis using different seeds}
	\label{table:seeds}
	
	\begin{tabular}{cccc}
		\toprule[1.5pt]
		
		Fuzzers   & \makecell[c] {\textbf{RSeed}}   & \textbf{PRseed} & \textbf{RPath} \\ \midrule
		AFLGo & 2311 & 52.4\% & 58  \\
		BEACON & 312 & 79.8\% &  23 \\
		WindRanger & 2532  & 64\%  & 41 \\   
		ParmeSan & 1463 & 43.8\% & 19 \\ 
            DeepGo   & 2788 & 72.6\% & 261\\ 
		\bottomrule[1.5pt]
	\end{tabular}
	\vspace{-1em}
\end{table}
From the results in Table \ref{table:seeds}, we can obtain two conclusions.
Firstly, by comparing the Rseeds and Rpaths of all fuzzers, we can see that DeepGo can generate more reachable seeds within the same time budget. Secondly, although BEACON has a higher Rpath, it has the lowest Rseed among all fuzzers due to pruning some reachable paths to the target site. Apart from BEACON, DeepGo has the highest PRseed among all fuzzers. \textbf{These three intermediate metrics suggest that DeepGo can generate optimal and viable paths to the target site by avoiding infeasible and hard-to-execute paths.}

\subsection{Case study of DeepGo's performance}
\label{case}

To show why DeepGo can reach the target sites faster, we used the target site \texttt{get\_audio.c}:1605 in \texttt{lame$_{3.99.5}$} as an example to conduct a case study. 
As listing \ref{lst1} shows, Line 24 is set as the target site. Since the path constraints at Line 2, Line 16, and Line 23 are all hard-to-satisfy path constraints, AFLGo, Windranger failed to reach the target site within the 24-hour time budget while DeepGo reached the target site (Line 24) at the 282$^{nd}$ minute.
As for AFLGo and WindRanger, AFLGo could reach Line 11-12, and WindRanger could reach Line 17. However, neither of them could satisfy the path constraints at Line 2, Line 9-10, Line 16, and Line 23 simultaneously to reach the target site. 
In Listing 1, conditions in different code locations are associated with different seed bytes (e.g., the values of $global.snd\_file$ in Line 2 and $global\_reader.input\_format$ in Line 11 are determined by different seed bytes) and mutating the relevant bytes would cause different path transitions and rewards. Since  AFLGo and WindRanger cannot predict the 
path transitions and rewards, they cannot avoid using actions that cause low-reward path transitions. Consequently, most of the new inputs generated by AFLGo and WindRanger could not cover Line 11-12 and Line 18, and thus failed to reach the target site (Line 24).

To identify such associations, during the testing process, the associations between the actions (mutation bytes) and the path transitions would be recorded. By querying the paths containing certain code locations, we can obtain the corresponding actions associated with them, thereby obtaining the correspondence between code locations and actions. Specifically, DeepGo can record the mappings between (path, action) and (next\_path, reward) to train VEE. If the path transitions caused by mutating a certain byte can guide DeepGo to cover interesting code locations (e.g., code locations closer to the target sites), the path transition and its corresponding action (mutation bytes) will be given a high reward. By querying the rewards of different path transitions, we can determine which bytes to mutate for interesting code locations.

\begin{figure}[b]
\centering
\setlength{\abovecaptionskip}{-0.1cm}
\includegraphics[width=\columnwidth]{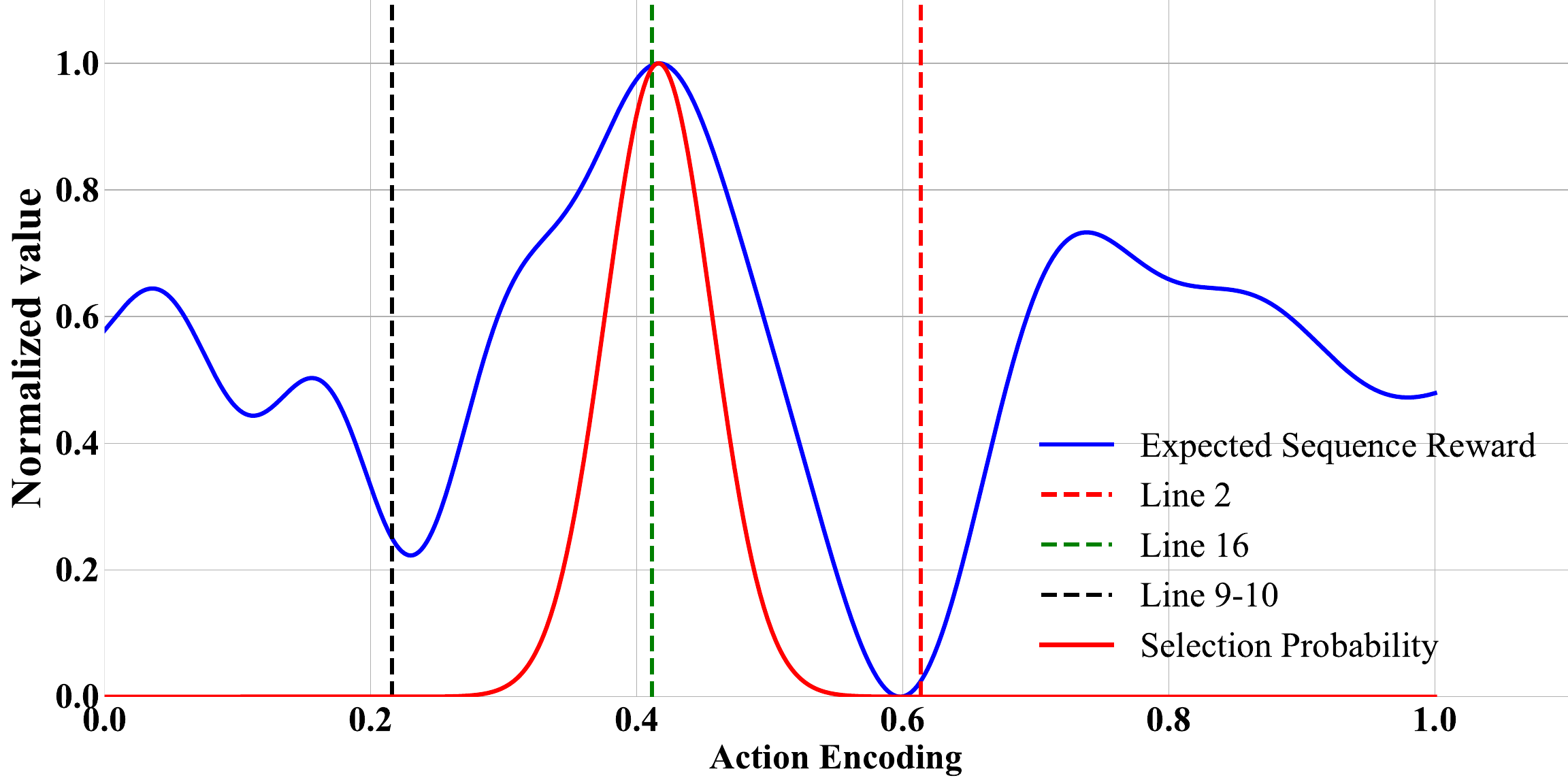} 
\caption{\label{fig:case} The expected sequence rewards and selection probability of actions.} 
\vspace{-1em}
\end{figure}
To prove the above analysis, we calculated the expected sequence rewards of taking all actions (i.e., mutating all bytes) and the selection probabilities of the RLF for each action after DeepGo reached Line 12 and presented the results in Fig. \ref{fig:case}.  In Fig. \ref{fig:case}, the x-axis 
represents the encoding of actions, and the y-axis represents the normalized value of the expected sequence
rewards and selection probabilities of actions. \texttt{Line 2}, \texttt{Line 9-10}, and \texttt{Line 16} represent three actions with encodings of 0.226, 0.41, and
0.618, respectively, which mutate the bytes related to the constraint condition in 
\begin{lstlisting}[caption={Example of the target site \textit{get\_audio.c:1605} in lame$_{3.99.5}$},captionpos=b,label=lst1,style=mystyle]
int init_infile(){
 if (global.snd_file == 0) {
  @global.music_in = open_wave_file(gfp, inPath);@
 }
 else
   ...
}
static FILE * open_wave_file(){
 if (global_reader.input_format != sf_raw && 
        (global_reader.input_format != sf_ogg){
  @global_reader.input_format = 
        parse_file_header(gfp,musicin);@
 }
}
static int parse_file_header(){
 if (type == IFF_ID_FORM) {
  @int const ret = parse_aiff_header(gfp, sf);@
 }
 else
   ...
}
static int parse_aiff_header(){
 if (dataType == IFF_ID_2CBE) {
  @global.pcmswapbytes = !global_reader.swapbytes;@
 }
 else
  ...
}
\end{lstlisting}
Line 2, Line 9-10, and Line 16, respectively. 
The blue line represents the distribution of the expected sequence rewards of actions, and the red line represents the Gaussian probability distribution of the RLF model for selecting actions. 
As the blue line shows, the normalized values of the expected sequence rewards of these three actions are 0.232, 0.994, and 0.061, respectively. It means that if the fuzzer takes actions at \texttt{Line 2} and \texttt{Line 9-10}, the caused path transitions  will make the fuzzer further away from the target site. 
As for the red line, the Gaussian probabilities of \texttt{Line 2} and \texttt{Line 9-10} are close to 0, while that of \texttt{Line 16} is close to 1. Thus, when selecting actions using Gaussian sampling, the RLF model tends to select the actions near \texttt{Line 16} rather than selecting 
\texttt{Line 2} and \texttt{Line 9-10}. By this means, DeepGo selects actions with high expected sequence rewards with a high probability, which helps avoid low-reward path transitions that go further away from the target site.

\section{Discussion}

\begin{figure}[b]
        \vspace{-0.2cm}
	\centering
	\setlength{\abovecaptionskip}{-0.1cm}
        \begin{adjustbox}{center=0cm}
	\includegraphics[width=7cm]{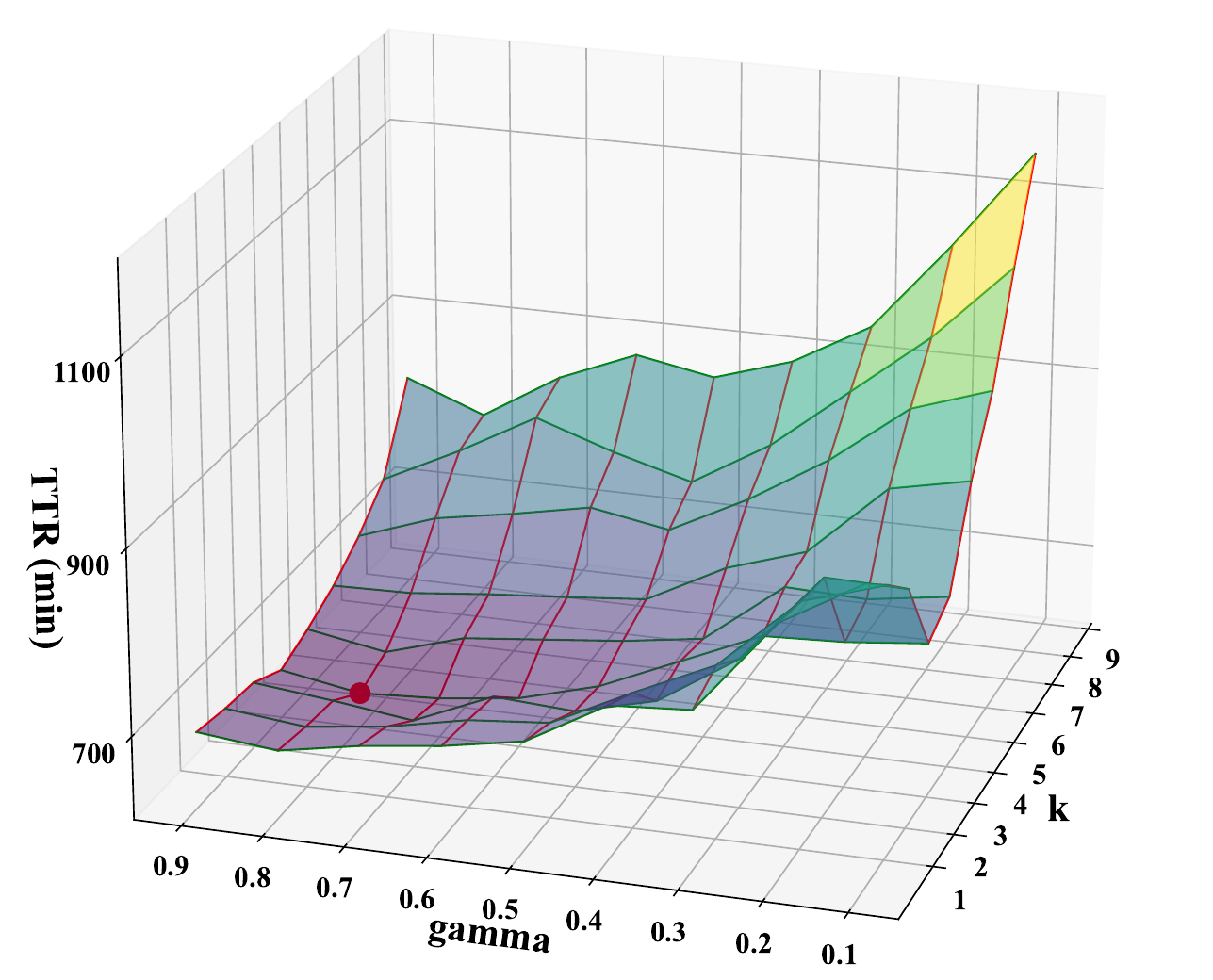} 
        \end{adjustbox}
	\caption{\label{fig:gamma}  The impact of hyperparameter settings on TTR.} 
\end{figure}

\textbf{Settings of hyperparameters.} 
The hyperparameter $\gamma$ and $k$ both affect DeepGo’s TTR.
 Firstly, in the training of the RLF model, the hyperparameter $\gamma$ is used to balance the influence of current rewards and subsequent rewards on the transition value and expected sequence rewards of the current state and its actions. Only focusing on current rewards, RLF may get stuck in local optima. On the contrary, overly emphasizing subsequent rewards may cause low-reward path transitions for the current path, thereby reducing the directed efficiency. Thus, the setting of the hyperparameter $\gamma$ would influence the Q-values, V-values, and policy of the RLF model, ultimately affecting DeepGo’s TTR. Secondly, in the $k$-step branch rollout strategy, the hyperparameter $k$ is used to generate the $k$-length path transition sequence. As $k$ increases, VEE is able to predict more path transitions, enabling RLF to have more foresight in designing policies. However, a high value of $k$ may lead to a decrease in the accuracy of VEE's predicted path transitions and rewards, misleading the RLF's policy to low-reward path transition sequences. Therefore, the setting of the hyperparameter $k$ would affect the prediction accuracy of VEE and the foresight of RLF's policy, ultimately impacting the TTR of DeepGo.
 
 To observe the impact of varying $\gamma$ and $k$ on DeepGo, we conducted experiments by setting $\gamma$ to 0.9, 0.8, 0.7, 0.6, 0.5, 0.4, 0.3, 0.2, and 0.1, and setting $k$ to 1, 2, 3, 4, 5, 6, 7, 8, and 9. We then utilized DeepGo with different hyperparameter configurations to test 20 programs from UniBench and recorded the mean TTR for each test case. In order to visually illustrate the impact of $\gamma$ and $k$ settings on DeepGo's TTR, we use a 3D chart to showcase the variations of TTR as $\gamma$ and $k$ change.
According to Fig. \ref{fig:gamma}, we can draw three conclusions. Firstly, the minimum TTR is achieved when $\gamma$ is set to 0.8 and the $k$ is set to 4, which is marked as a red point. Secondly, if the value of $\gamma$ is between [0.5, 0.9], and the value of $k$ is between [3, 5], the setting of $\gamma$ and $k$ has a relatively small impact on TTR (TTR changed within the range of [648, 688]).  Thirdly, TTRs are higher when $\gamma$ is greater than 0.5 compared to when $\gamma$ is less than 0.5, and TTRs are higher when $k$ is less than 5 compared to when $k$ is greater than 5.  This reflects the fact that since the length of path transition sequences in the fuzzing is generally less than 20 based on statistical fuzzing information, we should pay more attention to the influence of the current rewards of the path. Moreover, the setting of $k$ should balance the prediction and the foresight with an appropriate value. 

\textbf{Selection of targets.}
When selecting targets for the evaluation, we ran AFL++ for 48 hours because we believe that DGF techniques can faster reach the predefined targets than CGF techniques \cite{5, AFL++, ecofuzz, MobFuzzA}. 
Therefore, we believe that even if some targets are reached within more than 24 hours but less than 48 hours by CGF, they are still likely to be reached by DGF within 24 hours. 
For example, in our evaluations, out of the 51 targets that took AFL++ more than 24 hours to reach, 45 of them were reached by one or more directed fuzzers within 24 hours.
We use AFL++ rather than ``regular'' AFL to set the targets because AFL++ provides more comprehensive results, allowing for a detailed recording of the time required for CGF to reach different targets. With the information provided by AFL++, we can reproduce the code locations and the time cost to reach these locations, which is necessary for us to select the targets. In contrast, AFL does not provide such detailed information. 

\section{Related Work}
\textbf{Directed Symbolic Execution} (DSE) mostly relies on symbolic execution engines such as KLEE \cite{44}, KATCH \cite{KATCH}, and BugRedux \cite{38} to reach the target sites. Using program analysis and constraint solving, DSE can generate inputs that effectively penetrate through the path constraints toward the target sites. Although some state-of-the-art works, such as symcc \cite{40}, symqemu \cite{48}, and JigSaw \cite{49}, have been proposed to develop the symbolic execution, the heavyweight program analysis, path-explosion problem, and constraint solving of DSE still limit its scalability. 

\textbf{Directed Grey-box Fuzzing} (DGF) calculates the distances between the seeds and pre-defined targets to prioritize the seeds closer to the targets, which casts reachability as an optimation problem to minimize the distance between the seeds and their targets. Based on AFLGo's idea, Hawkeye \cite{21}, LOLLY \cite{60}, Berry \cite{berry}, UAFL \cite{58}, and CAFL \cite{22} proposed the new fitness metrics such as trace similarity, and sequence similarity, to enhance directedness and detect hard to manifest vulnerabilities. FuzzGuard \cite{27} filters out the unreachable seeds and BEACON \cite{26} prunes the infeasible paths, which are effective methods to improve the efficiency of DGF. 
$MC^{2}$ \cite{MC} models DGF as an oracle-guided search problem to find a target-reaching input, which accelerates the speed of reaching the targets. 
FISHFUZZ \cite{fishfuzz} enables fuzzers to seamlessly scale among thousands of targets and prioritize seeds toward interesting locations, thus achieving more comprehensive program testing.
SemFuzz \cite{semfuzz,23} analyzes the data-flow information and semantic information to generate valid input. 
Parmesan \cite{59}, V-Fuzz \cite{59}, and SAVIOR \cite{28} utilize the sanitizers, such as ASAN \cite{54} and UBSan \cite{Clang}, to label the potential buggy code as the target sites and steer the DGF to test the target sites. 
Hydiff \cite{36}, SAVIOR \cite{28} and Badger\cite{30} prioritize the seeds that may cause the specific program bug locations as the target sites, and then prioritize symbolic execution of the seeds that are reachable from more target sites. DrillerGO \cite{16} and Berry \cite{berry} combine the precision of DSE and the scalability of DGF to mitigate their individual weaknesses.
However, DGF still suffers from being difficult to penetrate through the hard-to-satisfy path constraints. DeepGo foresees critical execution information and predicts the optimal path. By combining the historical execution information and the predicted future execution information, DeepGo can  intelligently generate the optimal and viable path to the target site. By avoiding the infeasible and hard-to-execute paths, the fuzzer can reach the target site more precisely and efficiently.

\textbf{AI-Based Grey-box Fuzzing.} Previous state-of-the-art works \cite{learnfuzz, sky, HVLearn, NEUZZ, 27, Unit} apply AI techniques to augment the greybox fuzzing techniques. Among these works, NEUZZ \cite{NEUZZ} and MTFUZZ \cite{MTFUZZ} introduce gradient-descent based approaches to augment coverage-guided greybox fuzzing by approximating the PUT's discrete branching behavior. AthenaTest \cite{Unit}  uses the local transformer-based networks to extract features of seeds from the corpus and generate test cases. DYNFuzz \cite{DYNFuzz} and FuzzGuard \cite{27}  build models based on neural networks to predict whether the seeds are reachable to the target sites and filter out the unreachable seeds to enhance directedness. However, the existing AI-Based Grey-box Fuzzing techniques only optimize the selection of mutated bytes or seeds. They cannot comprehensively optimize all fuzzing strategies, making the fuzzer less intelligent in generating optimal paths to reach the target sites.

\section{Conclusion}
In this paper, we propose DeepGo, a predictive directed greybox fuzzer that can combine historical and predicted information
to steer DGF to reach the target site via an optimal path.
DeepGo constructs the Virtual Ensemble Environment, which uses DNNs to imitate the path transition model
and predict the rewards of potential path transitions. Using the RLF model, DeepGo combines the historical and predicted path transitions to determine the path transition sequence with the highest sequence rewards to generate optimal paths. Based on the MPSO algorithm, DeepGo optimizes the action group and exercises the high-reward path transition sequence to realize the optimal path. 
DeepGo is evaluated on 100 target sites of 25 real-world programs from 2 datasets, the experiment results show that DeepGo outperforms the state-of-the-art directed fuzzers (AFLGo, BEACON, WindRanger, and ParmeSan) in reaching target sites and exposing known vulnerabilities. Moreover, DeepGo also shows high accuracy in predicting the path transitions that have not been taken yet.

\section*{Acknowledgment}

This work is partially supported by the National Key Research and Development Program of China under Grant No. 2021YFB0300101, the National University of Defense Technology Research Project (ZK20-17, ZK20-09, ZK23-14), the National Natural Science Foundation China (62272472, 61902405, U22B2005, 61972412, 62306328), the HUNAN Province Natural Science Foundation (2021JJ40692), and the National High-level Personnel for Defense Technology Program (2017-JCJQ-ZQ-013).

\bibliographystyle{./IEEEtran}
\bibliography{./IEEEabrv,./IEEEexample}
\appendix

	\begin{table*}[hbp]
	\scriptsize
	\centering
	\setlength{\belowcaptionskip}{0.05cm}
	\setlength{\tabcolsep}{2pt}
	\caption{The TTR results on programs from UniBench}
	\label{table:unibench}
	
	\begin{tabular}{@{}ccccccccccc@{}}
		\toprule[1.5pt]
		\textbf{No} & \textbf{Prog} & \textbf{Version} &\textbf{Target sites} & {AFLGo}  & {BEACON} & {WindRanger} & {ParmeSan}  &{DeepGo-v} &{DeepGo-r} & {DeepGo} \\
		\midrule
		1  & \multirow{4}{*}{cflow} &    \multirow{4}{*}{1.6}  & parser.c:281 & T.O.  & 99.4m &61.1m & T.O. & 58.1m & T.O.& 41.8m \\
		2  &      &          & c.c:1783 & 12.8m  & 22.1m & 6.45m & 10.1m    & 12.2m & 16.4m & 11.2m \\
		3  &      &          &  parser.c:1223 &  1.22m  & 0.83m & 2.44m & 6.23m   &  6.16m &  4.76m &  8.23m\\
		4  &      &          & parser.c:108 & 12.8m  & 68.1m & 8.32m & 6.76m  &  13.5m & 17.2m &  16.1m \\ \hline
		
		5  & \multirow{4}{*}{mp42aac} &    \multirow{4}{*}{Bento4 1.5.1-628}      &   Ap4AvccAtom.cpp:82  &  T.O.  & N/A & T.O. &  N/A   & T.O. & T.O.& 891m \\
		6  &      &        &   Ap4TrunAtom.cpp:139 &  T.O. & N/A & T.O. &   N/A  & T.O.& T.O.& 723m  \\
		7  &      &        &  Ap4SbgpAtom.cpp:81&  T.O.   & N/A & T.O. &  N/A   & T.O.& T.O.& 1022m  \\
		8 &      &       & Ap4AtomFactory.cpp:490 &  T.O. &  N/A& T.O. &  N/A   & T.O.& T.O.& 878m \\ \hline
		9 & \multirow{4}{*}{jhead} &  \multirow{4}{*}{3.00}    &    exif.c:1339& T.O. & N/A & T.O. & T.O.  & T.O. & T.O.& 472m \\
		10 &      &       &   iptc.c:143  &  T.O. &N/A   & T.O. & T.O.   & T.O.& T.O.& 355m \\
		11 &      &       &    iptc.c:91 &  T.O. &N/A  & T.O. &  T.O. & T.O.  & T.O.& 892m \\
		12 &      &       &    makernote.c:174 &  T.O. &N/A  & T.O. &  T.O.  & 671m & T.O.& 98.1m \\ \hline
		13 & \multirow{4}{*}{mp3gain} &  \multirow{4}{*}{1.5.2}    &     layer3.c:1116 &  1142m  & N/A & 984m & N/A   & 871m & T.O. & 172m \\
		14 &      &       &   mp3gain.c:602 &  T.O.  & N/A & T.O. & N/A   &  T.O.& T.O.& 572m \\
		15 &      &      &    interface.c:663 &  T.O.  &  N/A& T.O. & N/A  &  T.O. & T.O.& 912m \\
		16 &      &      &   apetag.c:341 &  290m  & N/A & 91.2m  &  N/A &  164m &345m & 72.8m \\ \hline
		17 & \multirow{4}{*}{lame} &     \multirow{4}{*}{3.99.5}   &    bitstream.c:823 &   T.O.  &N/A  &T.O.  & N/A   &   T.O. &   T.O. & 521m \\
		18 &      &        &   lame.c:2148 &   T.O.  & N/A & T.O. &  N/A &   T.O.&   T.O. & 291m\\
		19 &      &        &  uantize\_pvt.c:441 &   T.O.  & N/A &  1269m  & N/A &   T.O. &   T.O.& 599m \\
		20 &      &         &   get\_audio.c:1605 &  T.O. &  N/A& T.O. &N/A &   932m &   T.O.& 412m \\ \hline
		21 & \multirow{4}{*}{imginfo} &  \multirow{4}{*}{jasper 2.0.12}  &    jp2\_cod.c:841 &T.O. &  N/A&   T.O. &  T.O. &  T.O. &   T.O.& 619m \\
		22 &      &         &  jp2\_cod.c:636 & T.O.  &  N/A & T.O. & T.O.  & T.O. &   T.O.& 776m \\
		23 &      &         & jas\_stream.c:823 & T.O.  &N/A   & T.O. &  T.O.  & T.O.&   T.O.& 351m\\
		24 &      &         &  jpc\_dec.c:1393& T.O.  & N/A &  T.O.  & 984m   & 641m &   T.O.&211m \\ \hline
		25 & \multirow{4}{*}{gdk-pixbuf-pixdata} & \multirow{4}{*}{gdk-pixbuf 2.31.1}  &    gdk-pixbuf-loader.c:387 & T.O.  & T.O. & T.O. & N/A & T.O. &   T.O.& 1126m  \\
		26 &      &         &   io-qtif.c:511  &T.O.   &  T.O. & T.O. & N/A & T.O. &   T.O.& 622m \\
		27 &      &         & io-jpeg.c:691 &  T.O.&  T.O. & T.O. & N/A & T.O. & T.O.& 498m  \\
		28 &      &         &  io-tga.c:360 & 126m  &  T.O. & T.O. &  N/A  & 87.3m & 172m &  48.7m \\ \hline
		29 & \multirow{4}{*}{jq} &   \multirow{4}{*}{1.5}  &    jv\_dtoa.c:3122 & T.O.  & T.O. & N/A & T.O.& T.O.& T.O. &1223m   \\
		30 &      &         &   jv\_dtoa.c:2004 & T.O.  & T.O. & N/A & T.O.& T.O. & T.O.& 1267m  \\
		31 &      &         & jv\_dtoa.c:2518 & T.O.  & T.O. & N/A & T.O.& T.O. & T.O.&  587m \\
		32 &      &         & jv\_unicode.c:42 & T.O.  &T.O.  &N/A  & T.O.  & T.O. & T.O.&  1024m \\ \hline
		33 & \multirow{4}{*}{tcpdump} &   \multirow{4}{*}{4.8.1}  &     print-aodv.c:259 & T.O.  & N/A & 843m  &T.O. & T.O. & T.O. & 622m  \\
		34 &      &       &     print-ntp.c:412 & 1436m  & N/A & 974m &1239m  & 1213m & T.O.& 542m  \\
		35 &      &        &     print-rsvp.c:1252  & T.O.  & N/A & T.O.  &889m & 338m & T.O.&  223m \\
		36 &      &         &     print-l2tp.c:606  & T.O.  & N/A & T.O.  &T.O. & T.O. & T.O.&  821m \\ \hline
		37 & \multirow{4}{*}{tic} &   \multirow{4}{*}{ncurses 6.1}    &     captoinfo.c:189 & T.O.  &N/A  & N/A & T.O. & T.O. & T.O.& 662m  \\
		38 &      &         &     alloc\_entry.c:141 & T.O. & N/A &  N/A&T.O.& T.O.  & T.O.&  761m \\
		39 &      &         &     name\_match.c:111 & 1186m  &N/A  &N/A  &T.O. & 251m & 1368m &  155m \\
		40 &      &          &     entries.c:78 &  1038m & N/A & N/A & 883m & 749m &    1411m& 495m \\ \hline
		41 & \multirow{4}{*}{flvmeta} &   \multirow{4}{*}{1.2.1}   &     json.c:1036 &T.O. & N/A & T.O. & T.O. & T.O. &T.O.& 682m   \\
		42 &      &         &     api.c:718 &T.O.   & N/A &  T.O.& T.O. & T.O. &T.O.&   577m \\
		43 &      &        &     flvmeta.c:1023 & T.O. & N/A &  T.O.&  T.O. & T.O.&T.O.&   1021m \\
		44 &      &         &     check.c:769 & T.O. &  N/A&  T.O.& T.O. & T.O. &T.O.&   874m \\ \hline
		45 & \multirow{4}{*}{tiffsplit} &   \multirow{4}{*}{libtiff 3.9.7}  &     tif\_ojpeg.c:1277 & T.O.  & N/A & T.O. &  N/A & T.O.&T.O.& 561m   \\
		46 &      &          &     tif\_read.c:335 &T.O.  &N/A  & T.O. & N/A  & T.O.&T.O.&  1123m  \\
		47 &      &         &     tif\_jbig.c:277 & T.O.  & N/A & T.O.  & N/A& T.O. &T.O.&  1046m  \\
		48 &      &         &     tif\_dirread.c:1977 & T.O. & N/A & T.O. & N/A  & 1068m &T.O. &  825m  \\ \hline
		49 & \multirow{4}{*}{nm} &  \multirow{4}{*}{binutils-5279478}  &     tekhex.c:325 & T.O.  & 364m  & 798m  & N/A& 369m &T.O. & 264m  \\
		50 &      &         &     elf.c:8793 & T.O.  & 986m & T.O. & N/A  & 665m &T.O. &  342m \\
		51 &      &         &     dwarf2.c:2378  & 1313m & 831m & 1065m &  N/A &  942m &T.O.& 212m  \\
		52 &      &         &     elf-properties.c:51 & T.O.  &T.O.  & T.O. &  N/A & T.O.& T.O. &  1062m \\ \hline
		53 & \multirow{4}{*}{pdftotext} &  \multirow{4}{*}{4.00}   &     XRef.cc:645  & T.O.  &  N/A & T.O. &  N/A & T.O.& T.O.& 427m  \\
		54 &      &         &     GfxFont.cc:1337 & 1345m &  N/A & T.O. & N/A & 1143m & 1423m&  785m \\
		55 &      &         &     Stream.cc:1004 &  725m &  N/A & T.O. & N/A & 498m & 911m&  352m \\
		56 &      &         &     GfxFont.cc:1643 &  637m &   N/A& T.O. &  N/A & T.O. & 876m &  T.O. \\ \hline
		57 & \multirow{4}{*}{sqlite3} &   \multirow{4}{*}{SQLite 3.8.9}   &     pager.c:5017 & 617m  & N/A & N/A & 1214m & 325m & 810m & 44.1m \\
		58 &      &         &     func.c:1029 & T.O.  & T.O. & N/A & T.O. & T.O. & T.O.&  1126m \\
		59 &      &         &     insert.c:1498 & T.O. & T.O. & N/A & T.O. & 310m  & T.O.&  452m \\
		60 &      &         &     vdbe.c:1984  & T.O. & 89.6m &N/A  & T.O. & T.O. & T.O.& 628m  \\ \hline
		61 & \multirow{4}{*}{exiv2} &   \multirow{4}{*}{0.26}    &     tiffcomposite.cpp:82 & 73.1m &N/A  & 68.1m & N/A&  57.2m & 93.2m & 42.1m  \\
		62 &      &        &     XMPMeta-Parse.cpp:847 & 37.5m  &N/A  & 21.4m &N/A  & T.O. & 79.5m &  T.O. \\
		63 &      &        &     tiffvisitor.cpp:1044 & 102m &N/A& T.O. &  N/A & 89.2m & 112m &  69.5m \\
		64 &      &         &    XMPMeta-Parse.cpp:896 & 86.7m  & N/A & 421m & N/A & T.O. & 99.1m &   T.O. \\ \hline
		65 & \multirow{4}{*}{objdump} &   \multirow{4}{*}{binutils-2.28}    &     elf.c:9509 &T.O.  & 782m  & T.O. & T.O. & T.O. & T.O.&  1294m  \\
		66 &      &        &     section.c:936 & T.O. & T.O.  & T.O. & T.O. & T.O. & T.O. &  1244m \\
		67 &      &        &     bfd.c:1108 & T.O. & 361m & 1288m &  T.O.& T.O. & T.O.&  1175m \\
		68 &      &         &    bfdio.c:262 & T.O.  & 1123m & T.O. & T.O. & T.O. & T.O.&  1032m \\ \hline
		69 & \multirow{4}{*}{ffmpeg} &   \multirow{4}{*}{4.0.1}    &     rawdec.c:268 & T.O. &N/A  & N/A & T.O. & 221m & T.O.&  183m  \\
		70 &      &        &    decode.c:557 & T.O.  &N/A  & N/A &N/A  & 338m & T.O.&  182m \\
		71 &      &        &     dump.c:632 & T.O. & N/A & N/A &  N/A & 673m & T.O.&  498m \\
		72 &      &         &    utils.c & T.O.  & N/A & N/A & N/A & 238m & T.O.&   178m \\ \hline
        73 & \multirow{4}{*}{mujs} &   \multirow{4}{*}{1.0.2}    &     jsrun.c:572 & T.O. &N/A  & T.O. & T.O. & T.O. & T.O.&  T.O.  \\
        74 &      &        &    jsgc.c:47 & T.O.  &N/A  & T.O. & T.O.  & T.O. & T.O.&  T.O. \\
        75 &      &        &     jsdump.c:292 & 532m & N/A & 421m & 433m & 488m & T.O.&  361m \\
        76 &      &         &    jsvalue.c:362 & T.O.  & N/A & T.O. & T.O. & T.O. & T.O.&   T.O. \\ \hline
            77 & \multirow{4}{*}{swftools} &   \multirow{4}{*}{0.9.2}    &     initcode.c:242 & 324m &N/A & 223m  & N/A & 298m & 401m &  196m  \\
		78 &      &        &    png.c:410 & 871m  &N/A  & 681m & N/A  & 501m & 1004m &  431m \\
		79 &      &        &     poly.c:137 & T.O. & N/A & T.O. &  N/A & T.O. & T.O.&  T.O. \\
		80 &      &         &    jpeg2swf.c:257 & 677m  & N/A & 541m & N/A & 611m & 881m &   481m \\

		\midrule
		\multicolumn{4}{c}{\fontsize{6}{5}\selectfont \textbf{speedup}}  & {\textbf{3.23$\times$}}  & {\textbf{1.72$\times$}} & {\textbf{1.81$\times$}} & {\textbf{4.83$\times$}}  &{\textbf{2.05$\times$}} &{\textbf{4.26$\times$}}& {\textbf{-}} \\ 
		\multicolumn{4}{c}{\fontsize{6}{5}\selectfont \textbf{mean $\hat{A}_{12}$}} & {\textbf{0.86}}  & {\textbf{0.81}} & {\textbf{0.83}} & {\textbf{0.89}} & {\textbf{0.83}}& {\textbf{0.90}} & {\textbf{-}} \\
		\multicolumn{4}{c}{\fontsize{6}{5}\selectfont \textbf{mean p-values}} & {\textbf{$0.008$}}  & {\textbf{$0.032$}} & {\textbf{$0.016$}} & {\textbf{$0.001$}}& {\textbf{$0.013$}} & {\textbf{$0.006$}} & {\textbf{-}} \\  
		\bottomrule[1.5pt]		
	\end{tabular}
         
\end{table*}
	
\end{document}